\documentclass[aps,twocolumn,nofootinbib,showpacs,floatfix]{revtex4}
\usepackage{graphics} 
\usepackage{amssymb,amsmath}
\usepackage{natbib}
\usepackage{amsmath}
\usepackage{psfrag}
\usepackage{graphicx}
\usepackage{subfig}
\usepackage{multirow}
\usepackage{color}
\usepackage{layout}
\usepackage{textcomp}
\usepackage{hyperref}
\usepackage[normalem]{ulem}

\def\al{\alpha} 
 
\def\de{\delta} 
 
\def\ep{\epsilon}
\def\ga{\gamma}

\def\be{\begin{equation}} 
\def\ee{\end{equation}} 
\def\bea{\begin{eqnarray}} 
\def\eea{\end{eqnarray}}  
\def\bean{\begin{eqnarray*}} 
\def\eean{\end{eqnarray*}}

\def\br{{\bf r}}

\def\bse{\begin{subequations}}
\def\ese{\end{subequations}}

\def\bV{{\mathbf V}}

\def\lsim{\raise 0.4ex\hbox{$<$}\kern -0.8em\lower 0.62ex\hbox{$\sim$}} 
\def\gsim{\raise 0.4ex\hbox{$>$}\kern -0.7em\lower 0.62ex\hbox{$\sim$}}

\def\f0N{f_0^{(N)}}
\def\bec{\begin{center}}
\def\eec{\end{center}}

\def\tv{{\text v}}

\begin{document} 
\title{{Classical particle scattering for  power-law two-body potentials}}

\author{ D. Chiron, B.~Marcos} 

\affiliation{Laboratoire J.-A. Dieudonn\'e, UMR 7351, Universit\'e de Nice --- Sophia Antipolis, Parc Valrose 06108 Nice Cedex 02, France} 

\begin{abstract}   
\begin{center}    
{\large\bf Abstract}\\

We present a rigorous study of the classical scattering for any
two-body inter-particle potential of the form $v(r)=g/r^\gamma$, with
$\gamma>0$, for repulsive ($g>0$) and attractive ($g<0$)
interactions. We give a derivation of the complete power series of the
deflection angle in terms of the impact factor for the weak scattering
regime (large impact factors) as well as the asymptotic expressions
for the hard scattering regime (small impact factors). We see a very
different qualitative and quantitative behavior depending whether the
interaction is repulsive or attractive.  In the latter case, the
families of trajectories depend also strongly on the value of
$\gamma$.  We also study carefully the modifications of the results
when a regularization is introduced in the potential at small scales.
We check and illustrate all the results with the exact integration of
the equations of motion.
\end{center}    

\end{abstract}    
\maketitle   
\date{today}  


\section{Introduction}

Scattering of particles are present in many physical processes in a
broad area of Physics, as atomic (e.g.~\cite{eisberg_61}), plasma
(e.g.~\cite{balescu_97}), astrophysics (e.g.~\cite{binney}), active
matter (e.g.~\cite{karamouzas_14}), etc. A seminal paper was published
by Ernest Rutherford in 1911 \cite{rutherford_11}, in which he studied
the deflection of $\alpha$ and $\beta$ particles by an atom.  He
calculated analytically the angle of deflection of the (positively
charged) incident particles with the (charged) nucleus. His
calculations, compared to experimental data (see \cite{rutherford_11}
for references), permitted to conclude that the atom is basically
``empty'' with a charge concentrated in the center, surrounded by the
electron cloud, which lead to the ``planetary'' model of the
atom. These two-body collisions plays also a central role in the
collisional relaxation of Coulomb plasmas (see e.g. \cite{balescu_97})
and self-gravitating systems (or more generally of systems of
particles with long range interactions), as pointed out by
Chandrasekhar in a seminal paper \cite{chandra_42}.  When studying the
relaxation of system of particles interacting with generalized
power-law interaction (see e.g. \cite{gabrielli_10,marcos_13}), it is
necessary to generalize the Chandrasekhar approach.

This paper is devoted to the rigorous mathematical study of the generalization of the classical scattering 
of two particles interacting with the generic power-law interaction
\be
\label{pot-def}
v(r)=\frac{g}{r^\gamma}.
\ee
Such process is well known only on the qualitative level or in
particular cases (see
e.g. \cite{eisberg_61,mcdaniel_64,drake_06,newton_02,kunc_97,friedrich_13}). For
example, in the case of a pure repulsive interaction, the angle of
deflection $\chi$ (defined in Fig.~\ref{coll}) is always well defined
for any value $\gamma$ and impact factor $b$ (see also
Fig.~\ref{coll}), and varies between $\chi=0$ (the particle comes back
in its original direction with opposite velocity) and $\chi=\pi$ (the
trajectory of the particle suffers no perturbation). In the case of
attractive interactions, the angle of deflection varies in the
interval $\chi\in[\pi,\infty[$. In this case two different situations
    arise: if $\gamma<2$, the angular momentum --- which scales with
    the distance as $1/r^2$ --- produces an effective repulsive
    interaction (the so-called centrifugal potential barrier) which
    always dominates the potential for $r\to0$ and the angle of
    deflection is finite for any $b$. However, for $\gamma>2$, the
    attractive interaction is stronger at small distances than the
    centrifugal barrier and particles can crash for values of the
    impact factor smaller than a critical quantity. For impact factors
    larger than this critical one, particles can make an arbitrarily
    large number of revolutions one around the other. This phenomenon
    is called in the literature {\it orbiting} (see
    e.g. \cite{friedrich_13} for a general discussion).

In this paper, we will consider interacting potentials of the form
\eqref{pot-def} with $\gamma>0$. The coupling constant $g$ is positive
for repulsive interactions and negative for attractive
ones. Analytical simple calculations are not possible except for some
particular cases for integer $\gamma$ in term of circular functions
(see \cite{whittaker_47}). In the other cases, only asymptotic
expansions can be performed. For this reason, we will derive the
asymptotic expressions for the angle of deflection of the particles
for the two limiting cases which are determined by the value of the
impact factor $b$ (defined in Fig.~\ref{coll}): the regime of {\it
  soft scattering}, in which the trajectories of the particles are
weakly perturbed, and the regime of {\it strong scattering}, in which
the particles suffer a large deflection. Moreover, we will study in
detail the introduction of a regularization (usually called {\it
  softening} in the astrophysical literature) at small scales in the
potential. This is of primarily interest when studying the relaxation
in systems of particles with long range interaction, in which the
effect of the regularization at small scales can play a key role (see
\cite{gabrielli_10}).

The paper is organized as follows: in the next section, we will review
definitions and standard formulas of the interaction of two particles
in a central force field. In the subsequent section, we will
explain the analytically tractable $\ga=1$ (Coulomb or gravitational)
case. Then, we will study mathematically the case $\ga\ne1$. We will
then first explain our general approach with the already known (see
e.g. \cite{mott-smith_60,landau1}) {\it soft scattering regime}, for
which we extend the domain of validity to arbitrary $\gamma>0$. Then,
we will derive expressions for the {\it hard scattering regime}, in
which we will obtain different classes of solution as a function of
$\gamma$. In the subsequent section we will explain the physical
implications of the mathematical results, compare them with the exact
numerical integration of the equation of motion and show typical
trajectories for the different regimes. Then, we will study how the
trajectories change when introducing a regularization at small scales
in the potential. We conclude the paper with a summary of the results,
conclusions and perspectives.

\section{Preliminaries}
\label{preliminaries}

Let us consider the scattering of two isolated particles. It is convenient to use the center of mass 
frame to transform the two-particle problem in a one-particle one. 
Let us consider that particles have masses $m_1$ and $m_2$ and their position $\br_1$
and $\br_2$ respectively. We define their relative position as
\be
\label{relative-vector}
\br=\br_1-\br_2
\ee
and fix the origin of the frame at the center of mass, i.e.,
\be
\label{center-of-mass}
m_1\br_1+m_2\br_2=\mathbf 0 .
\ee
The relation between the position of the particles in the center of
mass frame $\br$ and in the laboratory frame is, using
Eqs.~\eqref{relative-vector} and \eqref{center-of-mass}:
\bse
\label{change_coord}
\begin{align}
\br_1&=\frac{m}{m_1}\br\\
\br_2&=-\frac{m}{m_2}\br,
\end{align}
\ese
where we have defined the reduced mass
\be
m=\frac{m_1m_2}{m_1+m_2}.
 \ee
\begin{figure}
  \begin{center}
    \psfrag{C}{$\phi$}
    \psfrag{B}{$b$}
    \psfrag{A}{$\chi=2\phi$}
    \psfrag{D}{${\mathbf e}_{\parallel}$}
    \psfrag{E}{${\mathbf e}_{\perp}$}
    {\includegraphics[height= 0.25\textwidth]{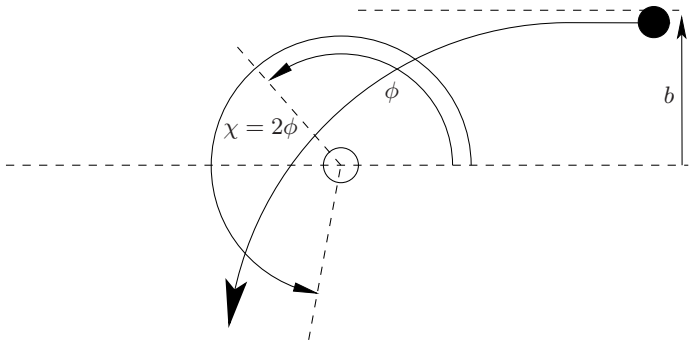}}
\caption{Collision in the center of mass frame. The black dot
  represents the fictitious (reduced) particle, and the white dot the
  center of mass of the particles, which is at rest.}
\label{coll}
  \end{center}
\end{figure}
In the center of mass frame, the collision occurs as depicted in Fig.~\ref{coll}, in which appears the 
definition of the impact factor $b$, the angle of closest approach $\phi$ and the angle of deflection $\chi$, 
which is $\chi=2\phi$. In order to define the angles with the usual mathematical signs, 
the incident particle comes from $+\infty$. This picture assumes that the two particles are far 
away from each other for $ t \to - \infty $ and for $ t \to + \infty $.
The angle $\phi$ can be calculated, as a function of the impact factor $b$, using the classical 
formula \cite{landau1}
\be
\label{phi}
\phi(b)=\int_{r_{min}}^{\infty} \frac{ (b/r^2) dr}{\sqrt{1 - (b/r)^2-2 v(r) / (mu^2) }},
\ee
where $u$ is the asymptotic velocity of the incident particle at $+\infty$ ($u=|\dot \br|$). 
The quantity $r_{min}$ is the largest positive root of the denominator, i.e., of
\be
\label{denominator}
W(r)=1 - (b/r)^2-2 v(r) / mu^2.
\ee
We consider the pure power law pair potential,
\be
v(r)=\frac{g}{r^\gamma},\qquad  0 <\gamma < d,
\ee
with $g \not =0$, where $g>0$  corresponds to a repulsive interaction and $g<0$ 
to an attractive one. We introduce the characteristic scale
\be
\label{b0gamma}
b_0=\left(\frac{|g|}{m u^2}\right)^{1/\gamma},
\ee
which allows us to rewrite Eq.~\eqref{phi} as
\be
\label{phi2}
\phi(b)=\int_{r_{min}}^{\infty} \frac{ (b/r^2) dr}{\sqrt{1 - (b/r)^2\mp2(b_0/r)^\gamma}}.
\ee
Now, the ``minus'' sign in the denominator corresponds to a repulsive  interaction while the ``plus'' sign 
to an attractive one. By using the change of variables $r=b/x$ it is possible to rewrite 
Eq.~\eqref{phi2} in the following form:
\be
\label{phi2-x}
\phi(b/b_0)=\int_0^{x_{max}} \frac{dx}{\sqrt{1 -x^2\mp 2(b_0/b)^\gamma x^\gamma}},
\ee
where $x_{max}$ is the smallest positive root of the
denominator. Since $x_{max} $ is a function of $ b/ b_0 $ depending
only on $ \ga $, Eq.~ \eqref{phi2-x} shows explicitly that $ \phi $ is
also a function of $ b/ b_0 $ depending only on $ \ga $. Equation
\eqref{phi2} can be solved analytically only in few cases (e.g.
gravity in $d=3$ which is given by $\gamma=1$), for the general case
approximations or numerical computation of the integral should be
used.

\section{$\gamma=1$ (Coulomb and gravitational case in $d=3$)}

In this section, we will first review the well-known Coulomb and gravitational case, which 
is analytically solvable. It will give us some insight for the general solution for $\gamma\ne1$.

We start from Eq.~\eqref{phi2}, compute the value of $r_{min}$ and the integral $ \phi $ explicitly. 
We obtain:
\begin{itemize}

\item for the repulsive case, $r_{min}=b_0+\sqrt{b^2+b_0^2}$ and
\be
\label{grav-rep}
\phi(b/b_0) = \arctan\left(\frac{b}{b_0}\right) ;
\ee
\item for the attractive case $r_{min}=-b_0+\sqrt{b^2+b_0^2}$ and
\be
\label{grav-att}
\phi(b/b_0) = \pi - \arctan\left(\frac{b}{b_0}\right) .
\ee
\end{itemize}
\begin{figure}
  \begin{center}
    \psfrag{X}{$b/b_0$}
    \psfrag{Y}[c]{{$\phi(b/b_0)$}}
   {\includegraphics[height= 0.3\textwidth]{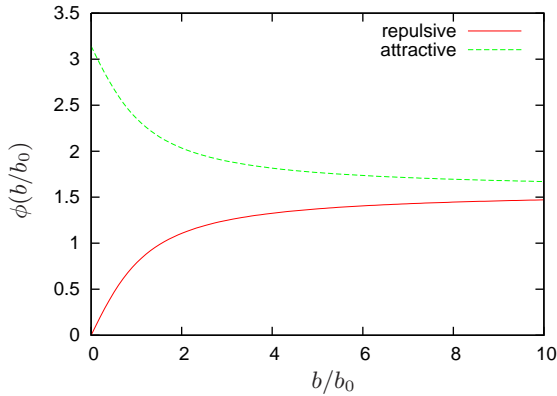}}
\caption{Graph of the angle $\phi$ as a function of $b/b_0$ for $\gamma=1$ and 
for repulsive (in red) and attractive (in green) interactions.}
\label{phi_grav}
  \end{center}
\end{figure}

We can identify two regimes in the collision process: the one corresponding to $b/b_0 \gg 1$, 
which is called  the ``weak'' or ``soft'' collisions regime, in which the trajectory is weakly 
perturbed; and the one corresponding to 
$b/b_0\ll 1$, which which is called the ``strong'' or ``hard'' collision regime, in which the trajectory is 
strongly modified by the collision.
From Eqs.~\eqref{grav-rep} and \eqref{grav-att} 
we obtain the following asymptotic behaviors for the angle $\phi$: 
\begin{itemize}
\item For the repulsive case, for weak collisions ($b/b_0\gg1$), we have 
$\phi(b/b_0)=\pi/2 -b_0/b +\mathcal{O} ( ( b_0 / b )^{3 } )$, and for strong ones 
($b / b_0 \ll 1) $, $\phi(b/b_0)=b/b_0+\mathcal{O} ( ( b / b_0 )^{3} )$.

\item For the attractive case, for weak collisions ($b/b_0\gg1$), we have 
$\phi(b/b_0)=\pi/2 +b_0/b +\mathcal{O} ( ( b_0 / b )^{3 } )$, 
and for strong ones ($b / b_0 \ll 1$), $\phi(b/b_0)=\pi -b/b_0+\mathcal{O} ( ( b / b_0 )^{3 } )$.
\end{itemize}

In the next sections we will compute the analogous asymptotic behaviors for the generalized 
case $\gamma\ne1$. 

\section{The general case: $\gamma\ne1$}
\label{section-general}

For the general case $\gamma\ne1$ it is not possible to derive an analytical expression for the 
angle $\phi$ as a function of $b/b_0$, as we did for $\gamma=1$ in Eqs.~\eqref{grav-rep} and 
\eqref{grav-att}. However, it is possible to compute the asymptotic behaviors 
of $\phi$ for  $b/b_0\ll1$ and $b/b_0\gg1$, which corresponds to hard and soft scattering respectively.

As a first step, we perform the substitution $ r = r_{min} / x $, $ 0 < x \le 1 $, in Eq.~\eqref{phi2}, 
yielding
\be
\label{substitu}
 \phi ( b / b_0 ) = \frac{b}{ r_{min} } \int_0^1 
 \frac{dx}{ \sqrt{1 - (bx / r_{min} )^2 \mp 2 (b_0 x/ r_{min})^\ga } } .
\ee
We recall  that the ``minus'' sign in the denominator 
corresponds to a repulsive  interaction while the ``plus'' sign to an attractive one. Then will use use the following procedure to compute the two limiting behaviors:
\begin{enumerate}

\item Determine, for the considered regime, an approximation for $r_{min}$ in Eq.~\eqref{substitu}, 
which is the largest zero of the denominator.

\item Perform an expansion in the appropriately chosen small parameter for each case 
of the denominator of Eq.~\eqref{substitu} and 
give an expression of the integrals by means of the $\Gamma $ function.

\end{enumerate}

We will study first the regime of soft collisions (i.e. $b/b_0\gg1$) for both attractive and 
repulsive interactions. Then, we will present in two different subsections 
(because the mathematical treatment is completely different), the case of hard scattering 
($b/b_0\ll1$) for repulsive interactions, and then for attractive ones.

\subsection{The regime of {\it soft} collisions for attractive and repulsive interactions}
\label{section-softcoll}
The regime of {\it soft} collisions corresponds to the case in which the scale $b_0$ is small 
compared to the impact factor $b$. In this regime the trajectories of the particles are weakly perturbed. In this Subsection, $\gamma $ is any positive number.

\begin{figure}
  \begin{center}
    \psfrag{X}{$r/b_0$}
    \psfrag{Y}{$W(r)$}
    \psfrag{AAAAAAAAA}{$\ga=1/2^-$}
    \psfrag{BBBBBBBBBB}{$\ga=1/2^+$}
    \psfrag{CCCCCCCCCC}{$\ga=1^-$}
    \psfrag{DDDDDDDD}{$\ga=1^+$}
    \psfrag{EEEEEEEE}{$\ga=3/2^-$}
    \psfrag{FFFFFFFF}{$\ga=3/2^+$}
    {\includegraphics[height= 0.35\textwidth]{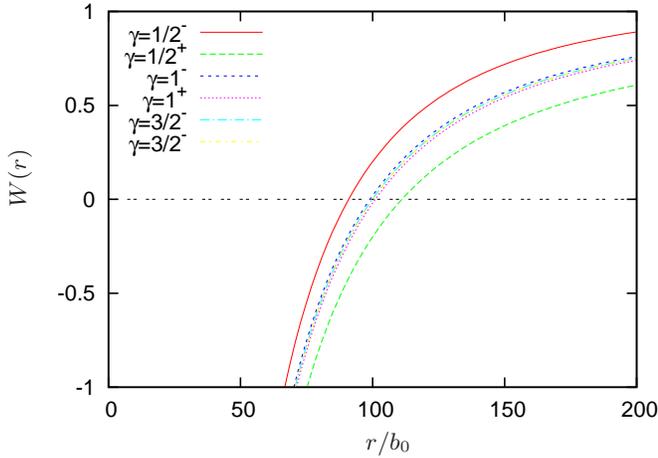}}
\caption{Graph of $W$ as a function of $ r/b_0$ 
for $b/b_0=100$ and different values of $\ga$ for the repulsive (superscript ``+'') 
and attractive case (superscript ``-''). 
Observe that in all cases $r_{min}\approx b$.}
\label{soft_case}
  \end{center}
\end{figure}

We first give an expansion of $ r_{min} $, which is the positive solution of $ 1 \mp 2 (b_0 / r_{min})^\ga =  ( b / r_{min} )^2 $. 
Recasting this as $ b^2 = r_{min}^2 \mp 2 b_0^\ga r_{min}^{2 - \ga } $, we see that if 
$ b \gg b_0 $, we must have $ r_{min} \gg b_0 $, hence 
$ b^2 = r_{min}^2 ( 1 \mp 2 ( b_0 / r_{min} )^\ga ) \approx r_{min}^2 $ 
(since $ \ga > 0 $) and then $ r_{min} \approx b \gg b_0 $. We see therefore that the 
value of $r_{min}$ does not depend, at leading order, on the sign of the interaction nor 
on the particular value of $\gamma$. This is illustrated in Fig.~\ref{soft_case}.

Expanding further yields
\begin{align}
\label{soft1}
 b / r_{min} & = \sqrt{ 1 \mp 2 ( b_0 / r_{min} )^\ga } 
 \nonumber \\ 
 & = 1 \mp ( b_0 / b )^\ga + \mathcal{O} ( ( b_0 / b )^{2 \ga} ) .
\end{align}
Next, we introduce the small parameter 
$ \de = 2 (b_0 / r_{min})^\ga = \mp [ ( b / r_{min} )^2 - 1 ] \approx 2 (b_0 / b )^\ga \ll 1 $ 
and obtain
$$
 \frac{ r_{min} }{ b } \phi (b/b_0) = \int_0^1 \frac{dx}{ \sqrt{1 - x^2 \mp \de ( x^\ga - x^2 ) }} .
$$
We want an expansion of the above integral using that $\de $ is a small parameter. It is then natural 
to write it under the form
$$
 \int_0^1 \frac{dx}{ \sqrt{1 - x^2 } \sqrt{1 \mp \de \frac{ x^\ga - x^2 }{1 - x^2 } }} 
$$
and to expand the second square root in power series. This is possible since the 
expression $ ( x^\ga - x^2 ) / ( 1 -x^2 ) $ is bounded on $ [0,1 ] $ (for $ \ga > 0 $) 
and this implies that $(r_{min} / b ) \phi (b/b_0 ) $ is actually a power series 
in $ \de $. In particular, we obtain the first order expansion 
\begin{align}
\frac{ r_{min} }{ b } \phi (b/b_0 ) & = 
\int_0^1 \frac{dx}{ \sqrt{1 - x^2} } 
\nonumber \\
& \quad \pm \frac{ \de }{ 2 } \int_0^1 \frac{x^\ga - x^2 }{ (1 - x^2)^{3/2}}\, dx + \mathcal{O} (\de^2 ).
\end{align}
Combining this with Eq.~\eqref{soft1} and using that $ \int_0^1 \frac{dx}{ \sqrt{1 - x^2} } = \pi /2 $ 
and that
\be
\label{integrale1}
 \int_0^1 \frac{1 - x^\ga }{ (1 - x^2)^{3/2}}\, dx = 
 A(\gamma)=\sqrt{\pi} \frac{ \Gamma \left( \frac{ \ga +1 }{2} \right) }{\Gamma \left( \frac{\ga}{2} \right)}
\ee
(see Appendix~\ref{app1}), we deduce
\begin{align}
\label{softcollisions}
 \phi ( b / b_0 ) = \frac{\pi}{2} \mp A(\ga) (b_0/ b )^{\ga} + \mathcal{O} ( (b_0/b)^{2\ga} ) .
\end{align}

On the mathematical level, we shall use the above strategy to give expansions with respect to 
some small parameter $ \de $ of integrals of the form 
\be
\label{prototype}
 \int_0^1 \frac{dx}{\sqrt{ F(x) + \de H(x) } } ,
\ee
where $F$ and $H$ are non-negative functions. If $ \int_0^1 \frac{dx}{\sqrt{ F(x)} } < +\infty $ 
and if $ H / F $ is bounded on $ [ 0, 1 ] $, then the above integral is an analytic function 
of $\de $ around $ \de = 0 $ and, for $ \de \to 0 $,
\begin{align*}
 \int_0^1 \frac{dx}{\sqrt{ F(x) + \de H(x) } } 
 & = \int_0^1 \frac{dx}{\sqrt{ F(x) } } 
 \\ 
& \quad - \frac{\de}{2} \int_0^1 \frac{H(x)}{F(x)^{3/2} } \, dx 
 + \mathcal{O}( \de^2 ). 
\end{align*}
If $ H(x) / F(x) $ is not bounded on $ [ 0, 1 ] $, then it may happen (and this 
is indeed true in some of the cases we shall study) that the above integral is not smooth with 
respect to $ \de $ and thus the correction is possibly not of order $\de $ 
but much larger.

The angle $ \phi $ is actually, for $b$ large enough, the sum of a power series in $ ( b_0 / b )^\gamma $, namely 
\be
\label{flowerpower}
 \phi ( b / b_0 ) = \sqrt{\pi} \sum_{n= 0}^{+\infty } \frac{ \Gamma ( (n \gamma +1) /2 ) }{ 2 n! \Gamma ( 1 + n ( \gamma /2 - 1 ) ) } ( \mp 2 ( b_0/ b)^\gamma )^n.
\ee
This formula has been established in \cite{mott-smith_60} for $ \gamma > 2 $ for both attractive and repulsive potentials, and converges for $ b > \beta b_0 $,
where 
\be
\label{b-definition}
\beta = \gamma^{1/\gamma} ( 1 - 2 / \gamma )^{\frac{2-\gamma}{2\gamma}}.
\ee

We have been able to extend (see Appendix \ref{appextension}) this formula for any $ \gamma > 0 $ and $ b > \beta b_0 $, where $ \beta ( \gamma=2 ) = \sqrt{2} $ and, for $ 0 < \gamma < 2 $,
$ \beta = \gamma^{1/\gamma} ( 2 / \gamma - 1 )^{\frac{2-\gamma}{2\gamma}} $.

\subsection{The regime of {\it hard} collisions for repulsive interactions}
\label{sect-hard-coll-repulsive}
\begin{figure}
  \begin{center}
    \psfrag{X}{$r/b_0$}
    \psfrag{Y}{$W(r)$}
    {\includegraphics[height= 0.35\textwidth]{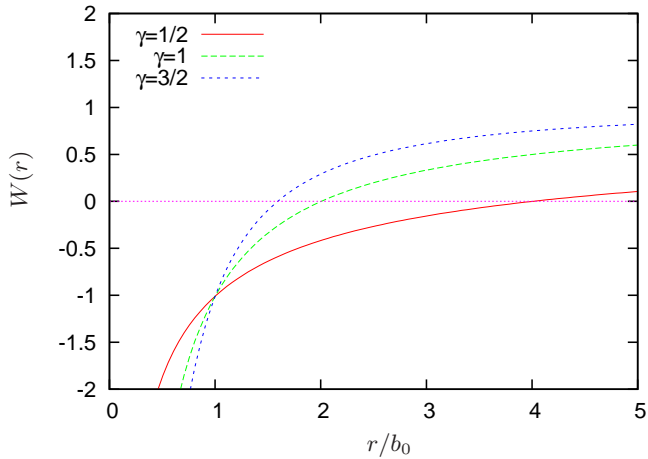}}
\caption{Graph of $W$ as a function of $ r/b_0$ 
for $b/b_0=1/10$ and different values of $\ga$ for the repulsive case. 
Observe that $r_{min}\sim b_0$.}
\label{hard_repulsive_case}
  \end{center}
\end{figure}

This corresponds to the minus sign in Eq.~\eqref{substitu}. In this Subsection again, 
$\gamma $ is any positive number. 
We first give the leading order of $ r_{min} $ by writing that 
$ 1 = ( b / r_{min} )^2 + 2 ( b_0 / r_{min} )^\ga $. Thus $ b \ll b_0 $ implies that 
$ r_{min} \to 2^{1/\ga} b_0 $. Then,
$$
 b / r_{min} \approx 2^{-1/\ga} b / b_0 ,
$$
and it follows that
\begin{align}
\label{hardmoins1}
 b / r_{min} = 2^{-1/\ga} b / b_0 + \mathcal{O} ( ( b / b_0 )^3 ).
\end{align}
In Fig.~\ref{hard_repulsive_case} we illustrate this behavior of $r_{min}$ by plotting $W$ for 
different values of $\ga$.
Here, the small parameter we consider is $ \delta = ( b / r_{min} )^2 \sim ( b / b_0 )^2 \ll 1 $ and 
substitute $ 2 ( b_0 / r_{min} )^\gamma = 1 - \de $ to obtain the expression 
$$
 \phi ( b / b_0 ) 
 = \sqrt{\de} \int_0^1 \frac{dx}{ \sqrt{ 1 - x^\ga + \delta ( x^\ga - x^2 ) } } ,
$$
which fits the form given in Eq.~\eqref{prototype}. Since the expression 
$( x^\ga - x^2 )/ ( 1 - x^\ga ) $ is bounded on $ [ 0,1 ] $, the above integral 
is here again a power series in $ \de $. In particular, we deduce the expansion
$$
\int_0^1 \frac{dx}{ \sqrt{ 1 - x^\ga + \de ( x^\ga - x^2 ) } } 
= \int_0^1 \frac{dx}{ \sqrt{ 1 - x^\ga } } + \mathcal{O} ( ( b/ b_0)^2 ) .
$$
Using the expression
\be
\label{integrale2}
\int_0^{1} \frac{dx}{\sqrt{1- x^\ga}} = \frac{ \sqrt{\pi } \Gamma\left(1+\frac{1}{\ga}\right)}{ 
 \Gamma\left(\frac{1}{2}+\frac{1}{\ga}\right)}
\ee
(see Appendix~\ref{app2}), we infer
\begin{align}
\label{durmoins}
 \phi ( b / b_0 ) = B(\ga) (b/ b_0 ) + \mathcal{O} ( (b/b_0)^{3} ) ,
\end{align}
where we have set
$$
B (\ga ) = \frac{2^{-1/\ga} \sqrt{\pi } \Gamma\left(1+\frac{1}{\ga}\right)}{ 
 \Gamma\left(\frac{1}{2}+\frac{1}{\ga}\right)} .
$$

\subsection{The regime of {\it hard} collisions for attractive interactions}
\label{sect-hard-coll-attractive}

We focus now on the plus sign in Eq.~\eqref{substitu} in the regime $ b \ll b_0 $. 
As we shall see, the situation is drastically different since the qualitative behavior 
strongly depends on $ \gamma $. We first give the leading order of $ r_{min} $ by writing that 
$ 1 + 2 ( b_0 / r_{min} )^\ga = ( b / r_{min} )^2 $. Thus, if $ b \ll b_0 $, 
we must have $ r_{min} \le b \ll b_0 $ and then 
$ 2 ( b_0 / r_{min} )^\ga \approx ( b / r_{min} )^2 $. Consequently, when $ \gamma \not = 2 $,
\begin{align}
\label{hardplus1}
 b / r_{min} \approx ( 2 b_0^\ga / b^\ga )^{ 1 / ( 2 - \ga ) } \gg 1 .
\end{align}
\begin{figure}
  \begin{center}
    \psfrag{X}{$r/b_0$}
    \psfrag{Y}{$W(r)$}
    {\includegraphics[height= 0.35\textwidth]{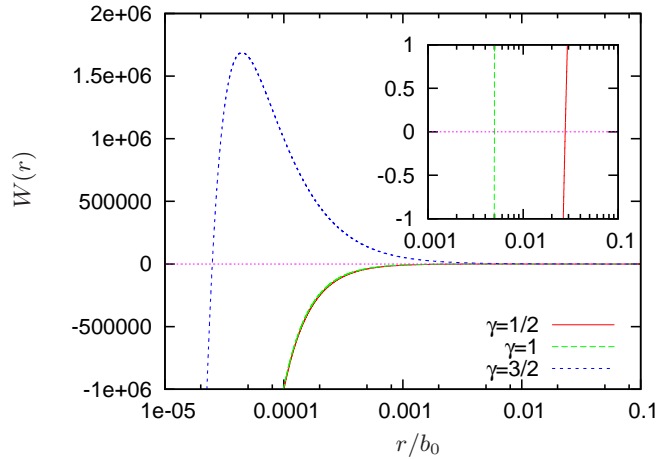}}
\caption{Graph of $W$ as a function of $ r / b_0 $ for $b/b_0=1/10$ and 
different values of $\ga$ for the attractive case. Observe that in this case $r_{min}\ll b_0$.}
\label{hard_case-att}
  \end{center}
\end{figure}

For this regime, we shall consider the small parameter $ \delta = ( r_{min} / b )^2 \ll 1 $ and 
substitute $ 2 ( b_0 / r_{min} )^\gamma = \delta^{-1} - 1 $ to obtain the expression
\be
\label{expresso}
 \phi ( b / b_0 ) = \int_0^1 \frac{dx}{ \sqrt{ x^\ga - x^2 + \de ( 1 - x^\ga ) } } ,
\ee
which tends, as $ \de \to 0 $, to $ \int_0^1 ( x^\ga - x^2 )^{-1/2} dx $, which is finite 
only for $ 0 < \ga < 2 $. 
This already leads us to study the case $ \gamma \geq 2 $ separately (see $\S$.~\ref{gamma-lager-2}). 
The expression on the right-hand side of Eq.~\eqref{expresso} fits the form in Eq.~\eqref{prototype}, 
but here, the situation is very different from the cases studied in Subsect.~\ref{section-softcoll} 
and \ref{sect-hard-coll-repulsive} since now, the expression $ ( 1 - x^\ga ) / ( x^\ga  - x^2 ) $ 
is {\it unbounded} on $ (0,1 ] $. Consequently, in the naive expansion of the right-hand 
side of Eq.~\eqref{expresso}
\begin{align*} 
& \int_0^1 \frac{dx}{ \sqrt{x^\ga -x^2} } 
 - \frac{\de}{2} \int_0^1 \frac{1 - x^\ga}{ (x^\ga -x^2)^{3/2} }  \, dx 
 \\ 
 & \quad + \frac{3 \de^2}{8} \int_0^1 \frac{ (1 - x^\ga)^2}{ (x^\ga -x^2)^{5/2} }  \, dx + \dotsc ,
\end{align*}
the first integral converges only for $ \ga < 2 $, the second one only for $ \ga < 2/3 $, 
the third one only for $ \ga < 2/5 $, etc. This suggests that on the one hand, $ \phi ( b / b_0 ) $ 
is probably not a power series in $ \de $ and on the other hand that we should separate the cases 
$ \ga < 2/3 $ (see $\S$~\ref{gammasmalldeuxtiers}) and $ 2/3 < \ga < 2 $ 
(see $\S$~\ref{gammadeuxtiersdeux}).

Before that, we may calculate, when $ 0 < \ga < 2 $, the leading order in $\delta$ of the 
integral Eq.~\eqref{expresso} 
\be
\label{hard-repulsive-zeroth-order}
\al(\ga)=  \int_0^1 ( x^\ga - x^2 )^{-1/2} \, dx 
 = \frac{\pi}{2- \ga },
\ee
We are going now to study the next order correction in the 
approximation of Eq.~\eqref{expresso} by Eq.~\eqref{hard-repulsive-zeroth-order}.

\subsubsection{$ 0< \gamma < 2/ 3 $}
\label{gammasmalldeuxtiers}

If $ \gamma < 2/ 3 $, the integral Eq.~\eqref{expresso} is indeed of class $ \mathcal{C}^1 $ with respect 
to $ \de $ (but probably not $\mathcal{C}^2$ when $ 2/5 < \ga < 2/3 $) and the differentiation under the 
integral sign is legitimated by the fact that 
$ \int_0^1 \frac{ 1 - x^\gamma }{ 2 ( x^\gamma - x^2 )^{3/2} } \, dx < \infty $. We then have
\begin{align*}
 \phi ( b / b_0 ) = \al(\ga) 
- \delta \int_0^1\frac{ 1 - x^\ga }{ 2 ( x^\ga - x^2 )^{3/2} } \, dx + o ( \delta ) .
\end{align*}
Reporting Eq.~\eqref{hardplus1} and using that
\be
\label{integrale3}
\int_0^1\frac{ 1 - x^\ga }{ 2 ( x^\ga - x^2 )^{3/2} } \, dx 
= \frac{ \ga}{ (2 -\ga)^2 } \frac{\sqrt{\pi} \Gamma \left(\frac{2 - 3 \ga}{2(2-\ga)} \right) }{ 
 \Gamma \left(\frac{2 (1- \ga)}{2-\ga} \right) } 
\ee
(see Appendix~\ref{app3}), we deduce
\begin{align}
\label{durplus1}
 \phi ( b / b_0 ) & = \al(\ga) 
 - C_1 (\ga ) ( b/ b_0 )^{ 2\ga / ( 2 - \ga ) } \nonumber \\
 & \quad + o ( ( b / b_0)^{ 2 \ga / ( 2 - \ga ) } ),
\end{align}
where we have defined
$$
C_1 (\ga) = \frac{ \ga}{ (2 -\ga)^2 } 2^{-2 /( 2-\ga) }  
 \frac{\sqrt{\pi} \Gamma \left(\frac{2 - 3 \ga}{2(2-\ga)} \right) }{ 
 \Gamma \left(\frac{2 (1- \ga)}{2-\ga} \right) } .
$$

\subsubsection{$ 2 /3 < \gamma < 2  $}
\label{gammadeuxtiersdeux}

We now assume $ 2 /3 \le \gamma < 2 $, for which Eq.~\eqref{expresso} is no longer expected to 
be of class $\mathcal{C}^1$ with respect to $\de $. We then write the correction 
$ \phi ( b / b_0 ) - \alpha (\gamma) $ under the form 
$ \phi ( b / b_0 ) - \alpha (\gamma) = - \de Q ( \de ) $, that is we define
\begin{align}
\label{bleue}
Q ( \de ) & = 
- \frac{ 1}{\de} \left( \phi(b / b_0) - \int_0^1 ( x^\ga - x^2 )^{-1/2} \, dx \right) \nonumber \\
& = \int_0^1 \psi_\de (x) \, dx ,
\end{align}
where we have set
$$ 
\psi_\de (x) = \frac{ ( x^\ga - x^2 )^{-1/2} ( x^\ga - x^2 + \de ( 1 - x^\ga ) )^{-1/2} (1-x^\ga) }{
\sqrt{ x^\ga - x^2 } + \sqrt{ x^\ga - x^2 + \de ( 1 - x^\ga ) } } .
$$
Clearly, as $ \de \to 0 $, $ Q ( \de ) $ tends to
$$ 
 \frac12 \int_0^1 \frac{ 1 - x^\ga }{ ( x^\ga - x^2 )^{3/2} } \, dx 
 = \int_0^1 \psi_0(x) \, dx = + \infty ,
$$
for $ \ga \ge 2 /3 $, due to the non integrable singularity at $ x= 0 $ (hence $ \phi $ is indeed 
not differentiable with respect to $\de $ at the origin). We then wish to determine the divergence 
speed in $ Q(\de ) $ as $ \de \to 0 $, and we shall show that actually $ Q( \de ) $ is of order 
$ \de^{1/\ga -3/2 } $ when $ 2/ 3 < \ga < 2 $ and of order $ \lvert \ln \de \rvert $ if $ \ga =2/3$. 

As a first step, we may get rid of the contribution for $ 1/2 \leq x \leq 1 $ in the integral 
of $ \psi_\de $ since
$$
 \int_{1/2}^1 \psi_\de (x) \, dx \to 
 \int_{1/2}^1 \frac{ 1 - x^\ga}{2 ( x^\ga - x^2 )^{3/2}} \, dx <  + \infty ,
$$
whereas $ Q(\de) \gg 1 $. Therefore,
$$
 Q(\de) = \int_0^{1/2} \psi_\de (x) \, dx + \mathcal{O}( 1 ) \approx \int_0^{1/2} \psi_\de (x) \, dx .
$$
Now, the idea is that the expression $ x^\ga - x^2 + \de ( 1 - x^\ga ) $ appearing in the 
denominator of $ \psi_\de $ is of order $ \delta $ if $ 0 \le x \le \de^{1/ \ga } $ and of order 
$ x^\ga - x^2 \sim x^\ga $ if $ \de^{1/\ga} \le x \le 1/2 $, which suggests to use the 
change of variable $ y = x / \de^{1/\ga} $ in the integral. Therefore,
\begin{align}
\label{mortel}
 Q ( \de ) & \approx \int_0^{1/2} \psi_\de (x)\, dx 
= \de^{ \frac{1}{\ga} - \frac{3}{2} } \int_0^{ \de^{- 1/\ga}/ 2 } \Psi_\de (y) \, dy ,
\end{align}
where we have set
\begin{align*}
& \Psi_\de (y) = ( 1 -  \de y^\gamma) \times \\ 
& \frac{ ( y^\ga - \de^{2/\ga -1} y^2 )^{-1/2} ( y^\ga - \de^{2/\ga -1} y^2 
+ 1 - \de y^\ga )^{-1/2}}{
\sqrt{ y^\ga - \de^{2/\ga -1} y^2 } + \sqrt{ y^\ga - \de^{2/\ga -1} y^2 
+ 1 - \de y^\ga } } . 
\end{align*}
As $ \de \to 0 $ and for $ 2/3 \leq \gamma < 2 $, one can justify rigorously that

\begin{align*}
 \int_0^{ \de^{- 1/\ga}/ 2 } \Psi_\de (y) \, dy \to & 
 \int_0^{ + \infty } \Psi_0 (y) \, dy 
\\ & = \int_0^{+ \infty} \frac{ y^{-\ga/2} ( y^\ga + 1 )^{-1/2}}{y^{\ga/2 } + \sqrt{ y^\ga + 1 } } \, dy,
\end{align*}
which is finite as soon as $ 2 /3 < \gamma < 2 $ since the integrand is $ \approx y^{ -3 \ga /2 } / 2 $ 
at infinity and $ \approx y^{ - \ga /2 } $ near the origin. We obtain finally, in Appendix~\ref{app4},
\be
\label{integrale4} 
 \int_0^{ + \infty } \Psi_0 (y) \, dy 
 = 
 \frac{2^{3 - 2 /\ga}}{\ga} \frac{ \Gamma \left( \frac32 - \frac{1}{\ga} \right) 
\Gamma \left( \frac{2}{\ga } -1 \right)}{\Gamma \left( \frac{1}{\ga } + \frac12 \right)} ,
\ee
and using Eq. \eqref{hardplus1} then gives, for $ 2/3 < \ga < 2 $ and $ b \ll b_0 $,
\begin{align}
\label{durplus3}
 \phi ( b / b_0 ) = \al(\ga) - C_3 ( \ga) ( b / b_0 ) + o ( b/b_0 ) ,
\end{align}
with
$$
 C_3 ( \ga) = \frac{2^{2- 3 /\ga}}{\ga} \frac{ \Gamma \left( \frac32 - \frac{1}{\ga} \right) 
\Gamma \left( \frac{2}{\ga } -1 \right)}{\Gamma \left( \frac{1}{\ga } + \frac12 \right)} .
$$

\subsubsection{ $ \ga = 2 /3 $}

It remains to study the case $ \ga = 2 /3 $, for which it is natural to expect from Eq.~\eqref{mortel} 
and the fact that
$$
 \Psi_0 (y) = \frac{1}{ y^{1/3} ( y^{1/3} + \sqrt{1 + y^{2/3} } ) \sqrt{ 1+ y^{2/3} } } 
 \approx \frac{1}{2y} 
$$
at infinity that 
\be
\label{guess}
 Q ( \de ) \approx \int_1^{ \de^{-3/2} / 2 } \frac{dy}{2y} 
 \simeq \frac34 \lvert \ln \de \rvert .
\ee
The mathematical justification of this result is given in Appendix \ref{app30}. 
Reporting this into Eq.~\eqref{bleue} and using that 
$ \de \approx b/ ( 2 \sqrt{2} b_0 ) $ by Eq.~\eqref{hardplus1} yields
\begin{align}
\label{durplus2}
 \phi ( b / b_0 ) = \frac{3 \pi}{ 4 } 
 - C_2 ( b/ b_0) \ln ( b_0 / b ) + o (( b/ b_0)\ln ( b_0/ b) ),
\end{align}
with $ C_2 = \dfrac{3}{ 8\sqrt{2} } $.

\subsubsection{$ \gamma = 2 $}
\label{gamma-egal-2}

The case $ \ga = 2 $ allows explicit computation and we see that it is a case where the attractive 
term is strong enough to form pairs when $b$ is small. Of course, this will be also the case 
when $ \gamma > 2 $. This implies that the function $ W $ in Eq.~\eqref{denominator} may have 
no positive zero. Actually, when $ \ga = 2 $, the behavior of the expression
$$
 W(r) = 1 - \frac{b^2}{ r^2} + 2 \frac{b_0^2}{r^2} = 1 - \frac{ b^2 - 2 b_0^2 }{ r^2 } 
$$
depends whether $ b > b_0 \sqrt{2} $ or $ b < b_0 \sqrt{2} $. If $ b > b_0 \sqrt{2} $, then 
$ W $ possesses $ r_{min} = \sqrt{ b^2 - 2 b_0^2 } $ as unique positive zero, and we have 
the exact value
\begin{align}
\label{phi-g-egal2}
 \phi ( b/ b_0 ) 
 & = \int_{r_{min}}^{+\infty } \frac{ (b/r^2) \, dr }{\sqrt{ 1 - r_{min}^2 / r^2 } } 
 \nonumber \\
 & = \frac{b \pi}{2 r_{min} } 
 = \frac{ \pi}{2 \sqrt{1 - 2 b_0^2 / b^2 } } .
\end{align}
If $ b \leq b_0 \sqrt{2} $, then $ W \geq 1 $ has no zero. This means that the two particles 
will crash one onto the other in finite time with a spiraling motion. The integral in the right-hand side 
of Eq.~\eqref{phi2} is then equal to $ + \infty $, but the angle $\phi$ has then no geometrical 
meaning and the picture given in Fig. \ref{coll} is then no longer the good one. There exists 
then a threshold $ b_0 \sqrt{2} $ with the property that particles crash  as soon 
as $ b \leq b_0 \sqrt{2} $.

\subsubsection{$\gamma > 2$}
\label{gamma-lager-2}
\begin{figure}
  \begin{center}
    \psfrag{X}{$r/b_0$}
    \psfrag{Y}{$W(r)$}
       {\includegraphics[height= 0.35\textwidth]{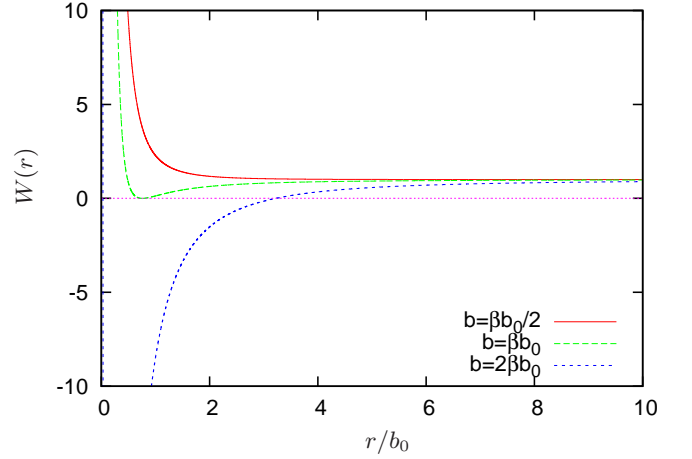}}
\caption{Graph of $W$ as a function of $r/b_0$ for different values of $b$ for $\ga=5/2$ 
for the attractive  case. Observe that for $b=\beta b_0/2$ there is no root, 
$b=\beta b_0$ is the limiting case with a double root and for $b=2\beta b_0$ there is one root.}
\label{hard_case-att-gammag2}
  \end{center}
\end{figure}

If $ \ga > 2 $, the attractive term is strong enough to form pairs for sufficiently small $b$, and 
we shall explicit the threshold. Notice first that when $ \ga > 2 $, the function 
$ W ( r) = 1 - b^2/r^2 + 2 b_0^\ga /r^\ga $ decreases 
on $ ( 0 , r_*(b) ] $ and increases on $ [r_*(b) , +\infty ) $, with
$$
 r_*(b) = \left( \frac{\ga b_0^\ga}{b^2 } \right)^{ \frac{1}{\ga -2} } .
$$
Since $ W ( r_*(b) ) = 1 - b^2/r_*^2(b) + 2 b_0^\ga/r_*^\ga(b) 
= 1 - (b_0/b)^{- \frac{2\ga}{\ga -2} } [ 1 - 2 / \ga ] \ga^{- \frac{2}{\ga -2}} $, 
we may then easily check that if
\be
\label{condition-binary}
b > \beta b_0,
\ee
where
\be
\label{defbeta}
\beta = \gamma^{1/\gamma}\left(1-\frac{2}{\gamma}\right)^{\frac{2-\gamma}{2\gamma}} ,
\ee
which coincides with the expression Eq.~\eqref{b-definition} appearing in the convergence radius of Eq.~\eqref{flowerpower}. Then $ W $ has a larger positive zero $ r_{min} $, whereas if $ b < \beta b_0 $, 
the expression $ W $ is positive on $ ( 0, +\infty ) $, and if $ b = \beta b_0 $, the expression 
$ W $ has a double root at $ r = r_* (\beta b_0) = b_0 ( \ga - 2 )^{1 / \ga } > 0 $, 
where $ W ( r_* (\beta b_0)) = 0 $. These three behaviors are illustrated in 
Fig.~\ref{hard_case-att-gammag2}.

When $ \gamma \to 2^+ $, we have, as expected, 
$ \beta = \gamma^{1/\gamma}\left(1- 2/\gamma \right)^{\frac{2-\gamma}{2\gamma}} 
= \gamma^{1/\gamma} \exp( (1/2) \left(1- 2/\gamma \right) \ln( 1- 2/\gamma ) ) \to \sqrt{2} $. 
If $ b < \beta b_0 $, the particles crash in finite time 
and $ \phi $ has here again no physical or geometrical meaning, despite the fact that the integral 
$$
 \int_0^{ +\infty } \frac{ (b/r^2) \, dr }{ \sqrt{ 1 - \frac{b^2}{r^2} + 2 \frac{b_0^\ga}{r^\ga } } } 
 = \int_0^{  + \infty } \frac{ dx }{ \sqrt{ 1 - x^2 + 2 ( b_0 / b )^\ga x^\ga } } ,
$$
where $ r_{min} $ has been replaced by $0$, converges. 

When $ b = \beta b_0 $, the reduced particle remain asymptotically trapped on a circular orbit 
of radius $ r_* (\beta b_0 ) > 0 $. This phenomenon is  called in the atomic physics literature 
{\it orbiting} (see e.g. \cite{friedrich_13}).
The angle $ \phi $ has once again no physical or geometrical meaning, and
$$
 \int_{r_*(\beta b_0 )}^{ +\infty } \frac{ (b/r^2) \, dr }{ \sqrt{ 1 - \frac{b^2}{r^2} + 2 \frac{b_0^\ga}{r^\ga } } } 
 = + \infty 
$$
in view of the fact that $ 1 - b^2 / r^2 + 2 b_0^\ga / r^\ga \sim ( r - r_*(\beta b_0 ))^2 $ 
for $r$ close to $ r_* (\beta b_0 )$.

Let us now consider the situation where we take $ \ga > 2 $ and $b$ slightly 
larger than $ \beta b_0 $, so that one expect a divergence in the integral $ \phi $. We have
$$
 \phi (b /b_0 ) = \int_{r_{min} }^{+\infty} \frac{b \, dr}{ r^2 \sqrt{ W_b (r) } } ,
$$
with $ W_b (r) = 1 - b^2/r^2 + 2 b_0^\ga /r^\ga $ (we have stressed the dependency 
on $b$ since we are interested in the limit $ b \to \beta b_0 $). 
We set $ R = r_*( \beta b_0 ) = b_0 ( \ga - 2 )^{1 / \ga } > 0 $. 
As $ b $ approaches $ \beta b_0 $, we have both $ r_*(b) \to R $ ($r_*(b) $ is the minimum for $ W_b $) 
and $ r_{min} \to R $ ($r_{min} $ is the largest zero of $W_b $). 
In the integral $ \phi $, the contributions for $r$ close to $ R $ will make the integral diverge 
since we shall have $ W_b (r) \sim ( r - R )^2 $ (we have a double root when $ b = \beta b_0 $), 
whereas the contributions for $ r $ much larger than $ R $ will remain of order one. 
As a consequence, for any small length parameter $ \ell > 0 $, we have
$$
 \phi (b /b_0 ) \approx \int_{r_{min} }^{ r_{min} + \ell } \frac{b \, dr}{ r^2 \sqrt{ W_b (r) } } ,
$$
and we may then replace $ W_b (r) $ by its second order Taylor expansion near $r_*(b)$: 
\begin{align*}
 W_b (r) & = W_b (r_*(b)) + ( r- r_*(b)) W_b' (r_*(b)) \\ 
 & \quad + \frac12 ( r- r_*(b))^2 W_b'' (r_*(b)) + \mathcal{O} ( (r-r_*(b))^3 ) .
\end{align*}
Since $ W_b'(r_*(b))= 0 $ and
\be
\label{deriveeseconde}
W_b''(r_*(b)) = \frac{2 \ga( \ga+1 ) b_0^\ga }{r_*^{\ga +2}(b) } 
- \frac{6b^2}{r_*^4(b)} \approx \frac{2 (\ga -2 ) b^2 }{ R^4} > 0 ,
\ee
this yields
\begin{align*}
 & \phi (b /b_0 ) \approx \int_{r_{min} }^{ r_{min} + \ell } 
 b r^{-2}\, dr /
 \\ & \sqrt{ W_b (r_*(b)) + ( r - r_*(b))^2 
 ( W_b''(r_*(b)) /2 + \mathcal{O} ( r - r_* (b)) ) } .
\end{align*}
We have $ W_b (r_*(b)) < 0 < W_b'' (r_*(b)) $ with $ W_b (r_*(b)) $ small but 
$ W_b'' (r_*(b)) $ of order one. The idea is then to use the substitution
$$ 
 z \sqrt{ - W_b (r_*(b)) } = ( r - r_*(b)) \sqrt{ W_b''(r_*(b)) /2 + \mathcal{O} ( r - r_*(b) ) } ,
$$
so that the expression in the square root in the integral becomes simply $ - W_b (r_*(b)) ( z^2 - 1 ) $. 
This yields
\begin{align}
\label{equivalent} 
\phi (b /b_0 ) \approx \frac{b}{ \sqrt{ - W_b (r_*(b)) } } \int_{1 }^{ z_{max} } 
 \frac{ r (z)^{-2} \, dr / dz }{  \sqrt{ z^2 - 1 } } \, dz ,
\end{align}
where $ z_{min} = 1 $ and $ z_{max} \approx Cte (\ell )  / \sqrt{ - W_b (r_*(b)) } \gg 1 $ are the 
corresponding values to $ r_{min} $ and $ r_{min} + \ell $. The idea is now that, roughly speaking, 
$ r (z) \approx r_*(b) \approx R $ and $ dr / dz \approx \sqrt{ - 2 W_b (r_*(b)) / W_b''(r_*(b))} $, 
which implies
\begin{align}
\label{integrale10}
 \phi (b /b_0 ) & 
 \approx \frac{b}{ R^2 } \sqrt{ \frac{2}{ W''_b (r_*(b)) } } \int_{1 }^{ z_{max}} \frac{ dz }{\sqrt{z^2 -1} } 
  \nonumber \\ 
 & \approx \sqrt{ \frac{ 2 b^2}{ R^{4} W_b''(R)} } \ln( z_{max} ) 
 \approx - \frac{\ln \lvert W_b(r_*(b)) \rvert }{ 2 \sqrt{ \ga - 2} } ,
\end{align}
in view of Eq. \eqref{deriveeseconde} and the fact that 
$ z_{max} \approx Cte (\ell ) / \sqrt{ - W_b (r_*(b)) } \gg 1 $. 
Finally, $ W_b(r_*(b)) = 1 - ( \beta b_0/ b)^{ -\frac{2\ga}{\ga -2} } $, and we end up with
\be
\label{phi-g-gt2}
 \phi (b /b_0 ) \approx - \frac{\ln( 1 - \beta b_0/ b )}{ 2 \sqrt{ \ga - 2} } .
\ee
For the sake of simplicity, we have included the mathematical details leading to Eq. \eqref{integrale10} 
in Appendix~\ref{app10}.

\section{Physical discussion}

In this section we  give a summary, a physical discussion 
and a numerical checking of the mathematical results derived in the previous section.

\subsection{Summary of the results and numerical checking}

We have summarized the results obtained in Sect.~\ref{section-general} in table \ref{yellowsummary}. 
In the first column, we make the difference between the regime of soft collisions, for which the 
angle $\phi\approx \pi/2$ (and hence the angle of deflection $\chi\approx \pi$) --- which means that 
the trajectories are weakly perturbed --- and the regime of strong collisions, in which the 
angle $\phi$ is far from $\pi/2$, i.e., the trajectory is strongly perturbed. In the case of 
attractive potentials, different cases arise depending on the value of $\ga $:
\begin{itemize}

\item For $0<\gamma < 2$, the leading order value of $\phi$ is $\pi/(2 -\gamma)$. 
The exponent of the first order correction depends 
whether $\gamma$ is smaller than $2/3$ or not. If we expand to higher order, we will see that the exponent of the second order term depends 
whether $\gamma$ is smaller than $2/5$ or not, the exponent of   the third order term whether $ \gamma$ is smaller than $2/7$ or not, and so on.

\item For $\gamma>2$, we have formation of pairs for impact factors smaller than a critical one. 
For impact factors exactly at the critical one there is the phenomena of {\it orbiting} and 
for larger impact factors the collision is well behaved.  We discuss these phenomena in the 
subsection below.

\end{itemize}

\begin{table*}
\begin{tabular}{| c || c || c |}
  \hline
 type of collision & repulsive potential & attractive potential  \\ 
   \hline \hline
  soft ($ b \gg b_0 $) 
  & $ \phi - \frac{\pi}{2} \sim - ( b_0 / b )^\ga $ 
  & $ \phi - \frac{\pi}{2} \sim + ( b_0 / b )^\ga $ \\
  \hline \hline
  hard ($ b \ll b_0 $) & $ \phi \sim b / b_0 $ & \begin{tabular}{ l | r }
$ \phi - \frac{\pi}{2 - \gamma } \sim - \left( \frac{b}{b_0} \right)^{2\ga/(2-\ga)} $ \quad & $ 0 < \ga < 2/ 3 $ 
\\ \hline 
$ \phi - \frac{ 3 \pi}{4} \sim - \frac{b}{b_0} \ln \left( \frac{b_0}{b} \right) $ \quad & $ \ga = 2/ 3 $
\\ \hline 
$ \phi - \frac{\pi}{2 - \gamma} \sim - \frac{b}{b_0} $ \quad & $ 2/3 < \ga < 2 $ 
\\ \hline
particles crash when $ b \leq \beta b_0 $ \quad & $ 2 = \ga $ 
\\ \hline 
\begin{tabular}{l}
particles crash when $ b < \beta b_0 $ 
\\ \hline 
formation of a binary (orbiting) when $ b = \beta b_0 $
\end{tabular} & $ 2 < \ga $ 
\end{tabular}\\ 
  \hline
\end{tabular}
\caption{Summary of the expansions of the angle $\phi $.}
\label{yellowsummary}
\end{table*}

\begin{figure}
  \begin{center}
  	\psfrag{AAAAAAAA}[c][c]{$\gamma=1/2$}
  	\psfrag{BBBBBBBB}[c][c]{$\gamma=1$}
  	\psfrag{CCCCCCCC}[c][c]{$\gamma=3/2$}
    \psfrag{X}{$b/b_0$}
         \psfrag{Y}[c][c]{$\pi/2-\phi(b/b_0)$}
        {\includegraphics[height= 0.35\textwidth]{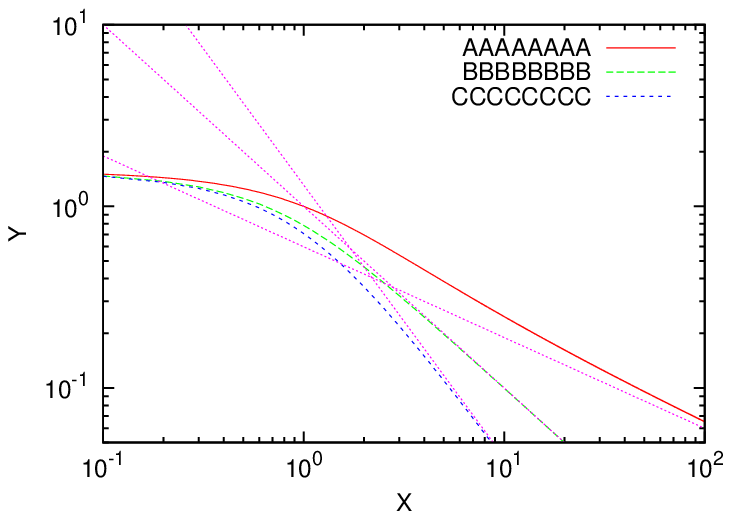}}\\
    \psfrag{Y}[c][c]{$\phi(b/b_0)$}
       {\includegraphics[height= 0.35\textwidth]{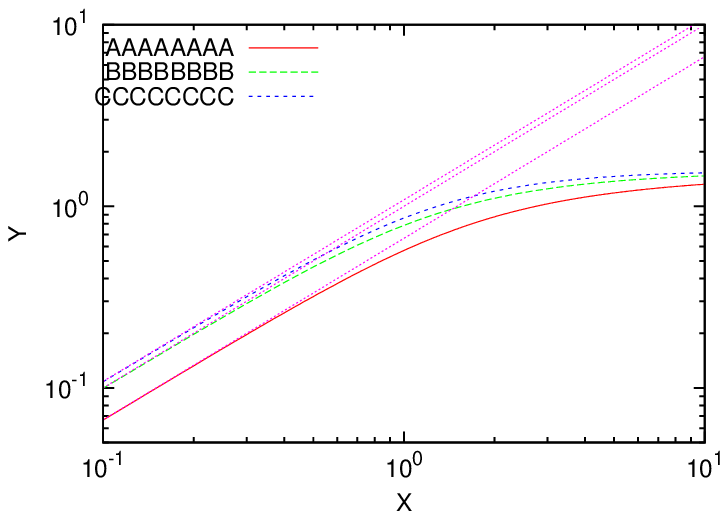}}
\caption{Numerical computations for repulsive potentials and several values of $\gamma$. 
Top: for soft scattering ($ b / b_0 \gg 1 $), plot of $\pi/2 - \phi$ (continuous line) 
and the theoretical predictions (dotted line) Eq.~\eqref{softcollisions} as a function of $b / b_0 $. 
Bottom: for hard scattering ($ b / b_0 \ll 1 $), plot of $ \phi$ (continuous line) and the 
theoretical predictions (dotted line) Eq.~\eqref{durmoins}  as a function of $b / b_0 $.}
\label{repulsive}
  \end{center}
\end{figure}

\begin{figure}
  \begin{center}
  	\psfrag{AAAAAAAA}[c][c]{$\gamma=1/2$}
  	\psfrag{BBBBBBBB}[c][c]{$\gamma=3/2$}
  	\psfrag{CCCCCCCC}[c][c]{$\gamma=2/3$}
    \psfrag{X}{$b/b_0$}
         \psfrag{Y}[c][c]{$\phi(b/b_0)-\pi/2$}
        {\includegraphics[height= 0.35\textwidth]{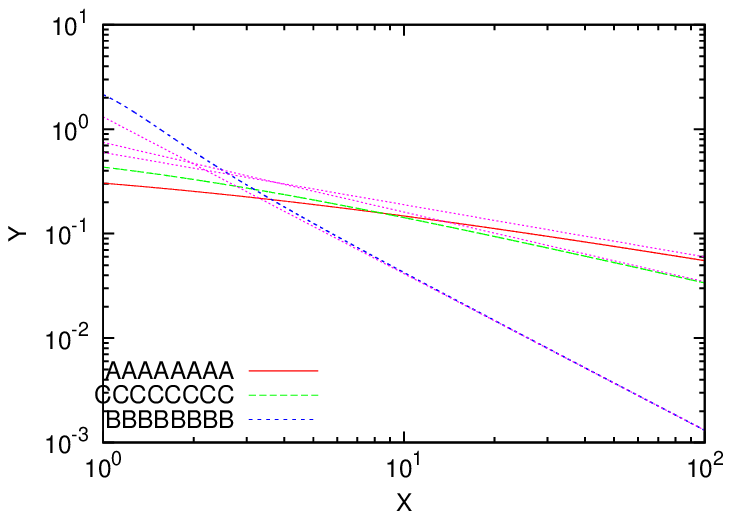}}\\
            \psfrag{Y}[c][c]{$\alpha(\ga)-\phi(b/b_0)$}
       {\includegraphics[height= 0.35\textwidth]{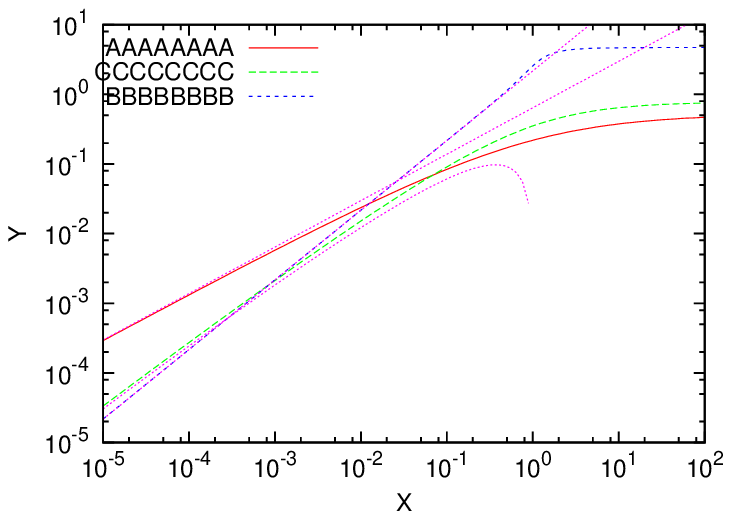}}
\caption{Numerical computations for attractive potentials and several values of $\gamma$. 
Top: for soft scattering ($ b / b_0 \gg 1 $), plot of $\pi/2 - \phi$ (continuous line) 
and the theoretical predictions (dotted line) 
Eq.~\eqref{softcollisions} as a function of $b / b_0 $. 
Bottom: for hard scattering ($ b / b_0 \ll 1 $), plot of $ \phi$ (continuous line) 
and the theoretical predictions (dotted line) 
Eq.~\eqref{durplus1}, \eqref{durplus3} or \eqref{durplus2} 
(depending on the value of $\ga$) as a function of $b / b_0 $.}
\label{attractive}
  \end{center}
\end{figure}

\begin{figure}
  \begin{center}
    \psfrag{X}{$b/b_0-\beta$}
    \psfrag{Y}[c][c]{$\phi(b/b_0)$}
   	{\includegraphics[height= 0.35\textwidth]{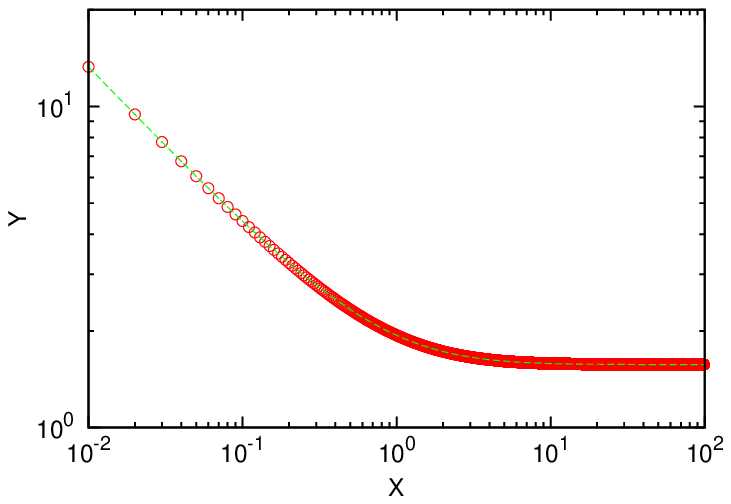}}\\
    {\includegraphics[height= 0.35\textwidth]{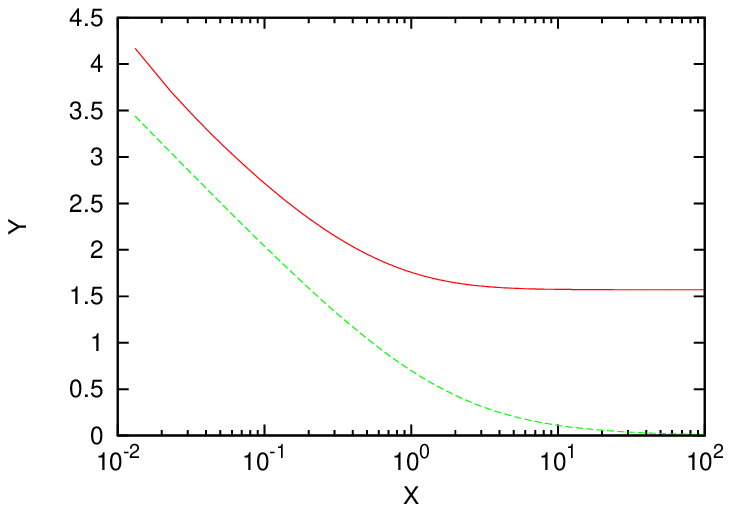}}
\caption{ Divergence of $\phi $ when $ 0 < b / b_0 - \beta \ll 1 $. 
Top: Graph of $\phi$ as a function of $b$ for $\gamma=2$ 
(continuous line) given by Eq.~\eqref{phi-g-egal2} and by the numerical calculation (circles). 
Bottom: same quantity for  $\ga=5/2$ (continuous red line) 
and the leading order given in Eq.~\eqref{phi-g-gt2} (dotted green line).}
\label{div-g-gt2}
  \end{center}
\end{figure}

In addition, for $ 0 < \ga < 2 $, we have checked numerically the validity of the asymptotic 
expansions in Sect.~\ref{section-general} for both repulsive potentials (Fig.~\ref{repulsive}) 
and attractive potentials (Fig.~\ref{attractive}), in the soft collision regime ($b / b_0 \gg 1 $) 
where the trajectory is weakly perturbed (top of each figure) and in the hard collision regime 
(bottom of each figure). We see a perfect matching between the numerical 
calculations and the analytical asymptotic calculations.
Finally, when $ \ga \geq 2 $, we illustrate in Fig.~\ref{div-g-gt2} the divergence 
of $\phi $ when $ b $ approaches $ \beta b_0 $ ($ b > \beta b_0 $) obtained in 
Eq.~\eqref{phi-g-egal2} for $\ga =2 $ (top) and in Eq.~\eqref{phi-g-gt2} when 
$\gamma=5/2 > 2 $ (bottom).

\subsection{Collisions with loops for attractive potentials}

Collisions with loops may appear when the angle $ \phi $ becomes large. This happens 
for attractive potentials in the two following cases: 
\begin{itemize}
 \item when $ \gamma $ is slightly smaller than $2$ and $ b \ll b_0 $, since then 
 $ \phi ( b /b_0 ) \approx \alpha ( \gamma ) = \frac{\pi}{2 - \gamma} $ (see 
 Subsect.~\ref{sect-hard-coll-attractive}).
 \item when $ \ga \geq 2 $ and $ b \approx \beta b_0 $ ($ b > \beta b_0 $), with the expression 
 for $\beta$ given in Eq.~\eqref{defbeta}, since (see $\S$ ~\ref{gamma-lager-2})
\bse
\label{summary-betagt2}
\begin{align}
\phi\left(\frac{b}{b_0}\right)&=\frac{\pi}{2\sqrt{1-2b_0^2/b^2}}
\quad \quad \mbox{if } \ga = 2 , \\
\phi\left(\frac{b}{b_0}\right)&\approx-\frac{\ln(1-\beta b_0/b)}{2\sqrt{\ga-2}} 
\quad \quad \mbox{if } \ga > 2 .
\end{align}
\ese
\end{itemize}

It is interesting to study these trajectories, for which it is numerically 
convenient to use for the first part of the trajectory the implicit relation between the polar angle $\theta\in[0,\phi]$  and the 
distance to the origin $r$ of the particle (see e.g. \cite{landau1}):
\be
\label{phi-func-r}
\theta(b,b_0,r)=\int_{r}^{\infty} \frac{ (b/r'^2) dr'}{\sqrt{1 - (b/r')^2- 2 ( b_0/r')^\ga }}.
\ee
Note that $\theta(b,b_0,r_{min})=\phi(b/b_0)$.
The first half of the trajectory is therefore
\bse
\begin{align*}
x&=r \cos\theta\\
y&=r \sin\theta,
\end{align*}
\ese
and the second one
\bse
\begin{align*}
x&=r \cos(2\phi-\theta)\\
y&=r \sin(2\phi-\theta).
\end{align*}
\ese
We see that the trajectory is symmetric about a straight line which passes by the origin of coordinates 
(i.e. the center of mass) and the point of closest approach (defined by the angle $\phi$). In the plot, 
the first half part of the trajectory --- from $x= + \infty$ to the axis of symmetry  --- is 
plotted in red, the other half of the trajectory in green. 
The points of intersection of the trajectory lie on the axis of symmetry.

\subsubsection{The case $ \ga = 2^- < 2 $ and $ b / b_0 \ll 1 $}

In Subsect.~\ref{sect-hard-coll-attractive} we have seen that, for the attractive potential 
with $\gamma<2$, we have
\be
\label{limit-phi}
\lim_{b/b_0\to0} \phi\left(\frac{b}{b_0}\right) = \alpha(\ga) = \frac{\pi}{2-\ga}
\ee
and (see the beginning of that Subsect.)
$$
 r_{min} \leq b \ll b_0 .
$$
Therefore, in the limit $ \ga \to 2$, the angle $ \phi ( 0^+ ) $ diverges. 
Fixing $ \ga < 2 $ but $ \ga \approx 2 $ (say $\ga = 1.95 $ for instance), we have, 
for $ b / b_0 \ll 1 $, a collision where $ r_{min} \ll b_0 $  is 
very small and 
$ \phi $ very large, which corresponds to many loops in a very small region close to 
the center of mass of the two particles. 

\begin{figure}
  \begin{center}
  	\psfrag{AAAAAAAAA}[c][c]{$\gamma=1/2$}
  	\psfrag{BBBBBBBB}[c][c]{$\gamma=1$}
  	\psfrag{CCCCCCCC}[c][c]{$\gamma=3/2$}
    \psfrag{X}{}
    \psfrag{Y}{}
       {\includegraphics[height= 0.35\textwidth]{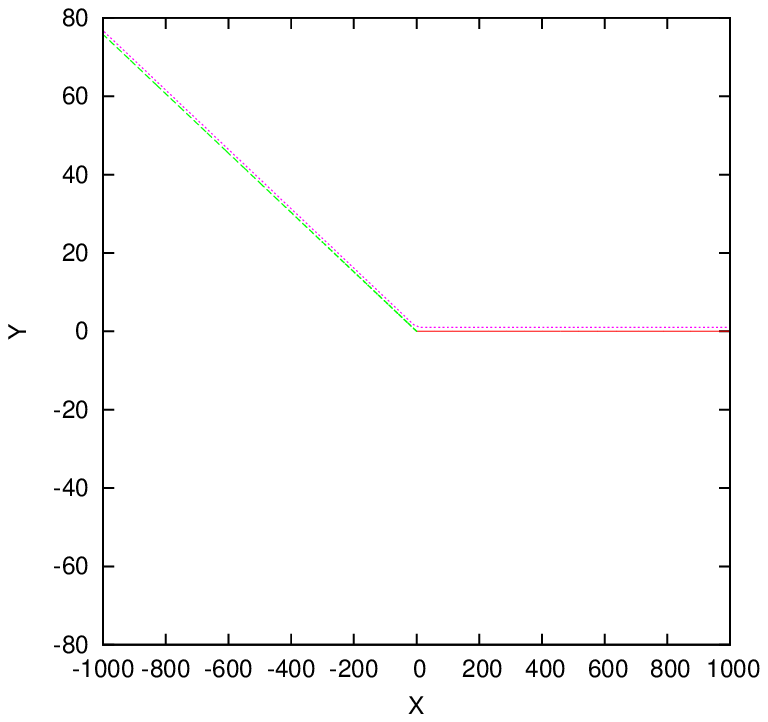}}\\
         \psfrag{Y}{}
        {\includegraphics[height= 0.35\textwidth]{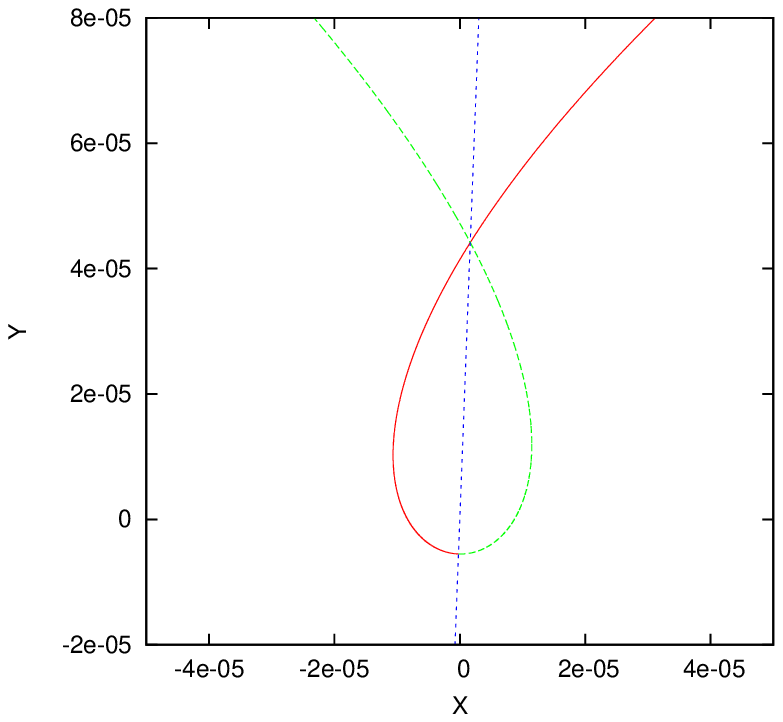}}
\caption{Top: A trajectory in the center of mass frame for attractive hard interaction and $\ga=4/3$ and $b/b_0=0.025$. The pink dotted line represents the prediction of Eq.~\eqref{durplus3} (with a small offset). Bottom: zoom of the plot above, 
in which a loop is visible. The first half part of the trajectory --- from $x=+\infty$ to 
the axis of symmetry  --- is plotted in red, the other half of the trajectory in green. 
The points of intersection of the trajectory lie on the axis of symmetry.}
\label{loop}
  \end{center}
\end{figure}

\begin{figure*}
  \begin{center}
   \psfrag{X}{}
    \psfrag{Y}{}
    \psfrag{A}{$\phi$}
    \psfrag{B}{$b$}
    \psfrag{C}{$\theta$}
    \psfrag{D}{${\mathbf e}_{\parallel}$}
    \psfrag{E}{${\mathbf e}_{\perp}$}
    {\includegraphics[height= 0.3\textwidth]{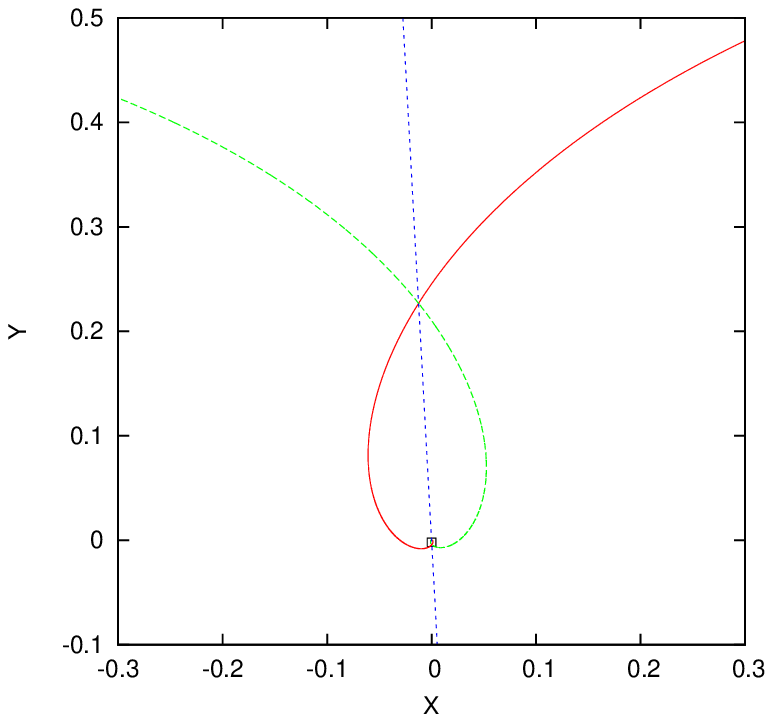}}
     {\includegraphics[height= 0.3\textwidth]{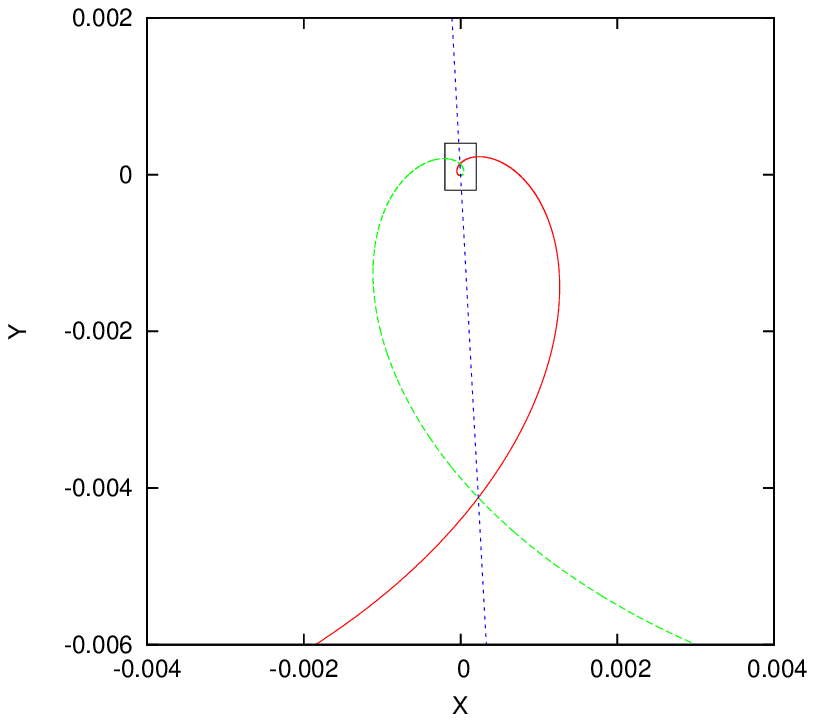}}
     {\includegraphics[height= 0.3\textwidth]{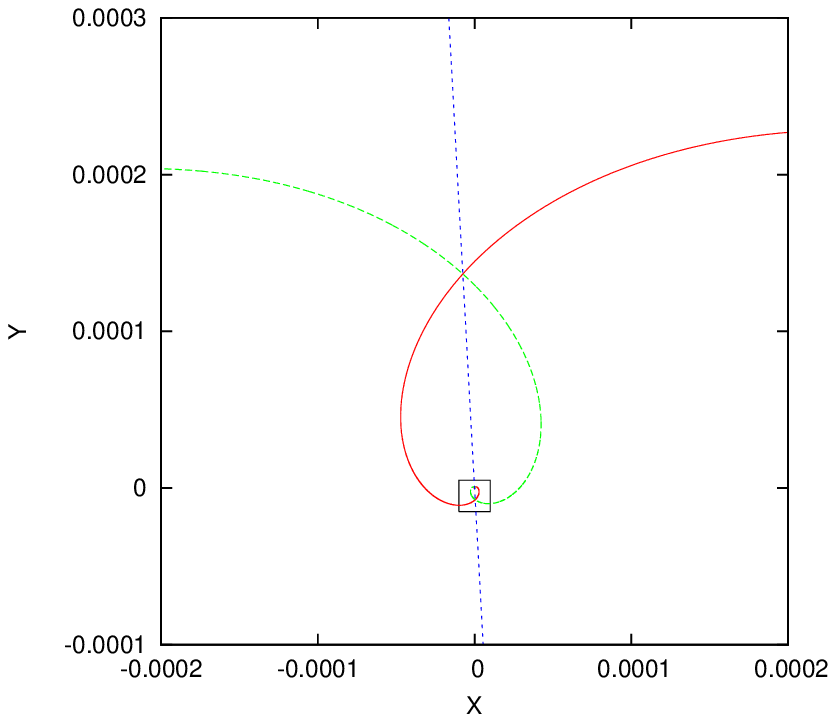}}\\
      {\includegraphics[height= 0.3\textwidth]{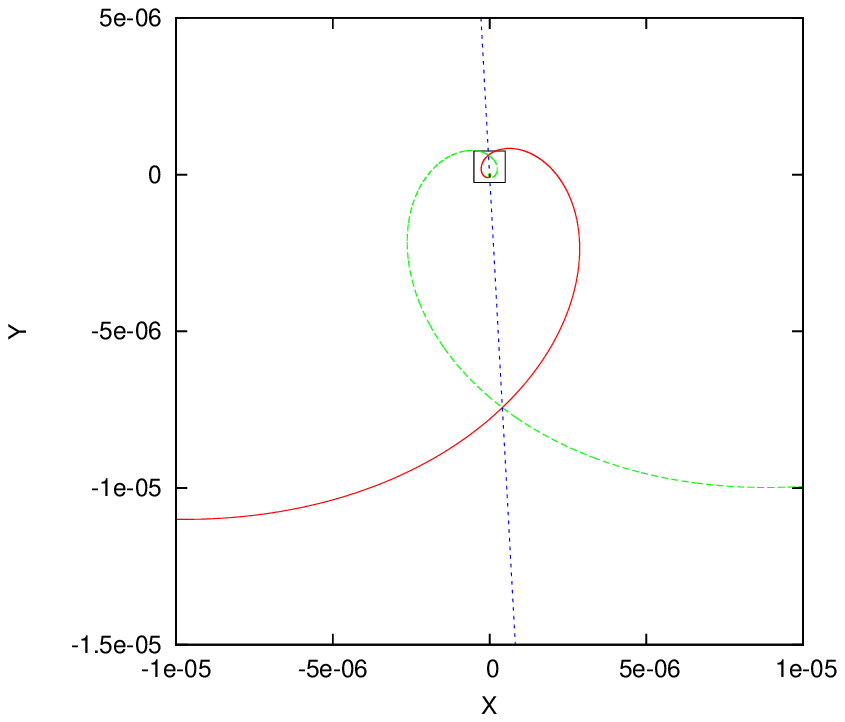}}
     {\includegraphics[height= 0.3\textwidth]{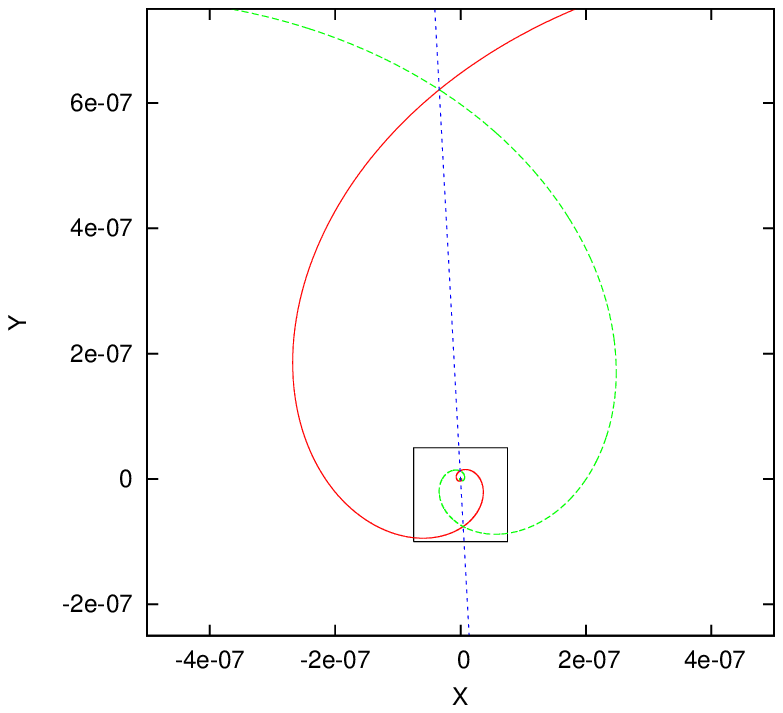}}
     {\includegraphics[height= 0.3\textwidth]{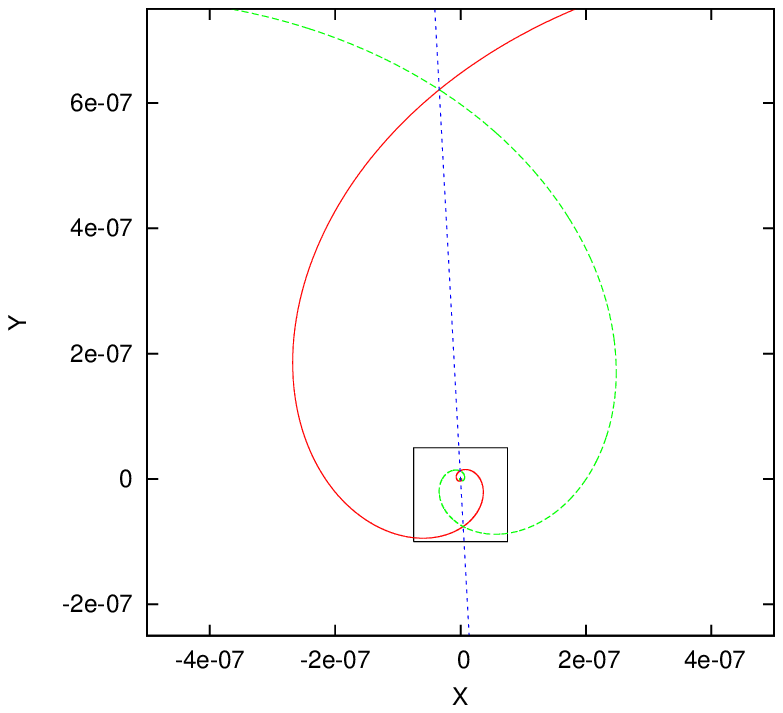}}\\
      {\includegraphics[height= 0.3\textwidth]{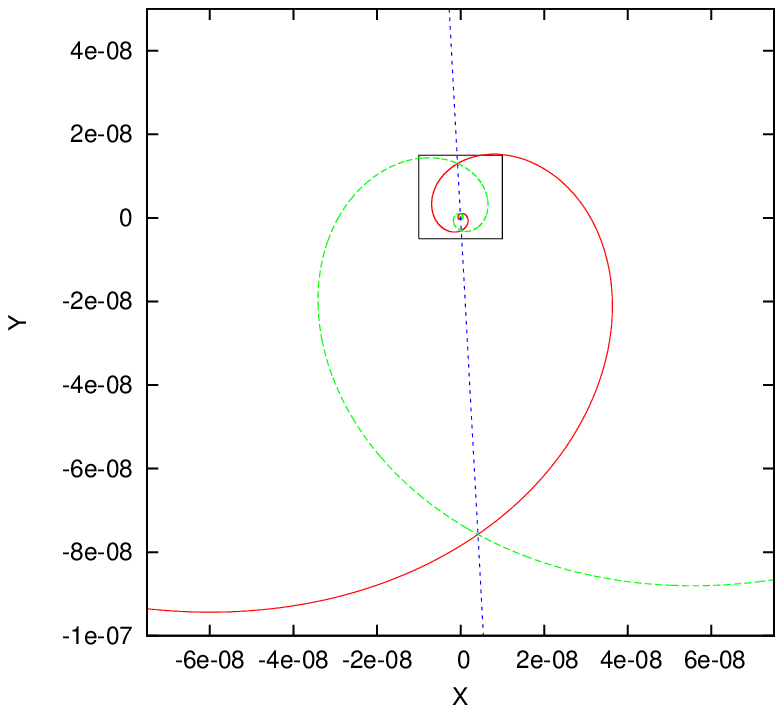}}
     {\includegraphics[height= 0.3\textwidth]{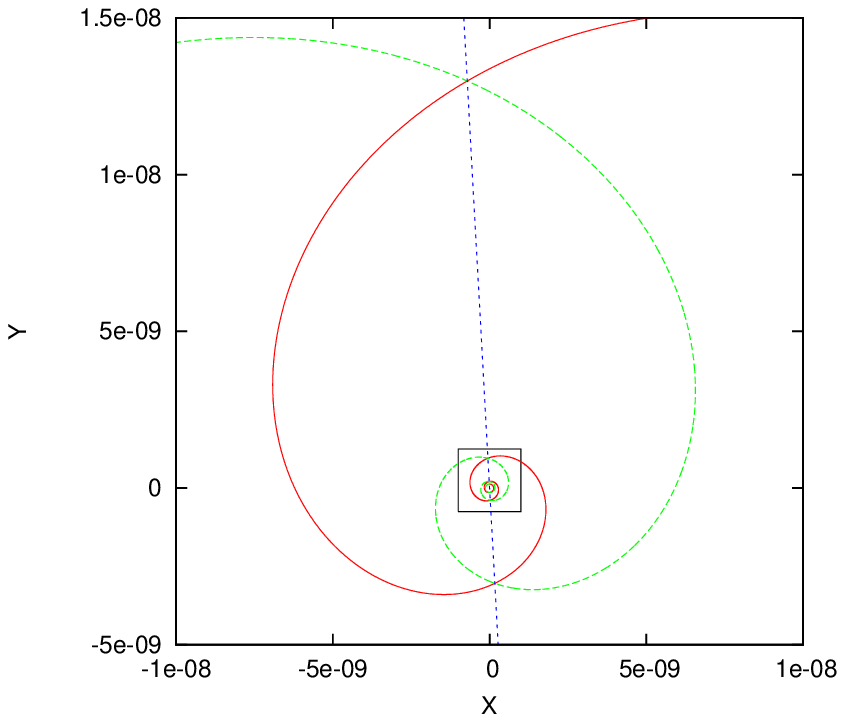}}
     {\includegraphics[height= 0.3\textwidth]{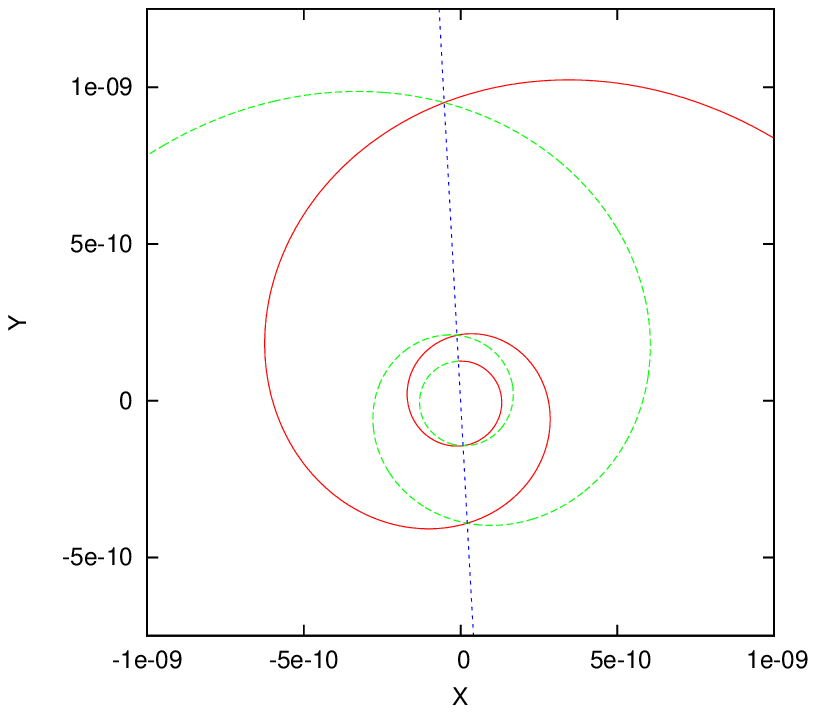}}
\caption{Collision in the center of mass frame for $\gamma=1.95$ and $b/b_0=0.8$. The dotted line is 
the axis of symmetry of the trajectory. The square in each plot represents the frame of the next plot 
(which have to be read from left to right and top to down). The first half part of the 
trajectory --- from $x=+ \infty$ to the axis of symmetry  --- is plotted in red, the other 
half of the trajectory in green. 
The points of intersection of the trajectory lie on the axis of symmetry.}
\label{coll2}
  \end{center}
\end{figure*}

The number of intersections between the first half of the trajectories (red lines in the plots) and the 
symmetric one (green lines in the plots) depends on the value of $\phi$. The 
number of intersections of the trajectories (and hence the number of loops) is given by the floor 
function of the angle of closest approach divided by $\pi$
\be
\label{nloops}
n_{loops}\left(\frac{b}{b_0}\right)=\mathrm{floor}\left(\frac{\alpha(\gamma)}{\pi}\right) 
= \mathrm{floor}\left(\frac{1}{2 - \gamma}\right) .
\ee
In the example showed in Fig.~\ref{loop}, $\gamma=4/3$, $\phi\approx 4.67$, and hence there is one loop. 
A more complex example appears when choosing $\ga=1.95$ and $b/b_0=0.8$. 
In this case $\phi\approx 39.3$ and hence the numbers of loops is, using Eq.~\eqref{nloops}, 
$n_{loops}=12$. This can be shown explicitly in Fig.~\ref{coll2}, where we show successive 
zooms in the trajectory, in which appear smaller and smaller loops.

\begin{figure}
  \begin{center}
    \psfrag{X}{$r/b_0$}
    \psfrag{Y}{$\theta(b,b_0, r )$}
       {\includegraphics[height= 0.35\textwidth]{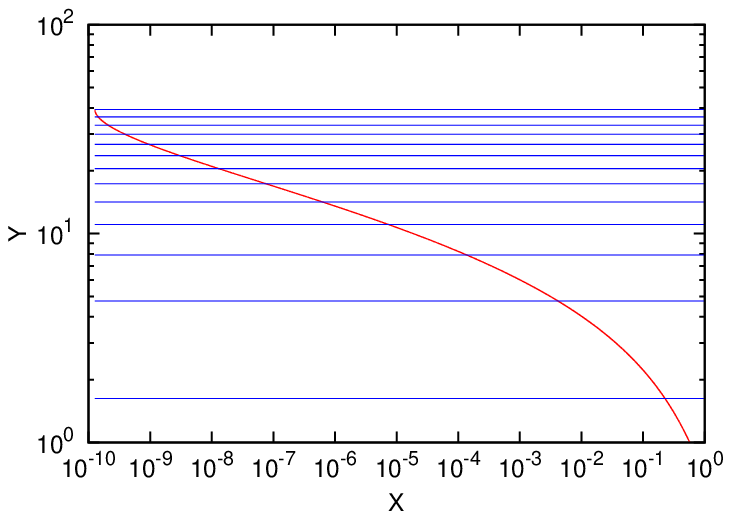}}
\caption{Graph of $\theta(b,b_0, r )$ as a function $r/b_0$ for a 
trajectory with $\ga=1.95$ and $b/b_0=0.8$. 
Each horizontal line corresponds to an intersection in the trajectory.}
\label{loop2}
  \end{center}
\end{figure}

In Fig.~\ref{loop2}, we show $\theta( b,b_0 , r ) $ (in Eq.~\eqref{phi-func-r}) as a function of $r$. 
Each horizontal line corresponds to $ \phi = n\pi$, in which $n$ is an integer and with $n$ 
such that $\phi(b/b_0)> n\pi$. 

\subsubsection{The case $ \ga \geq 2 $ and $ b / b_0 = \beta^+ $}

We have seen in Subsect.~\ref{gamma-lager-2} that the angle $\phi$ diverges for $b/b_0\approx \beta$, 
with the expression for $\beta$ given in Eq.~\eqref{defbeta}, see the formulas recalled in 
Eq.~\eqref{summary-betagt2}.

\begin{figure}
  \begin{center}
    \psfrag{X}{}
    \psfrag{Y}{}
        {\includegraphics[height= 0.35\textwidth]{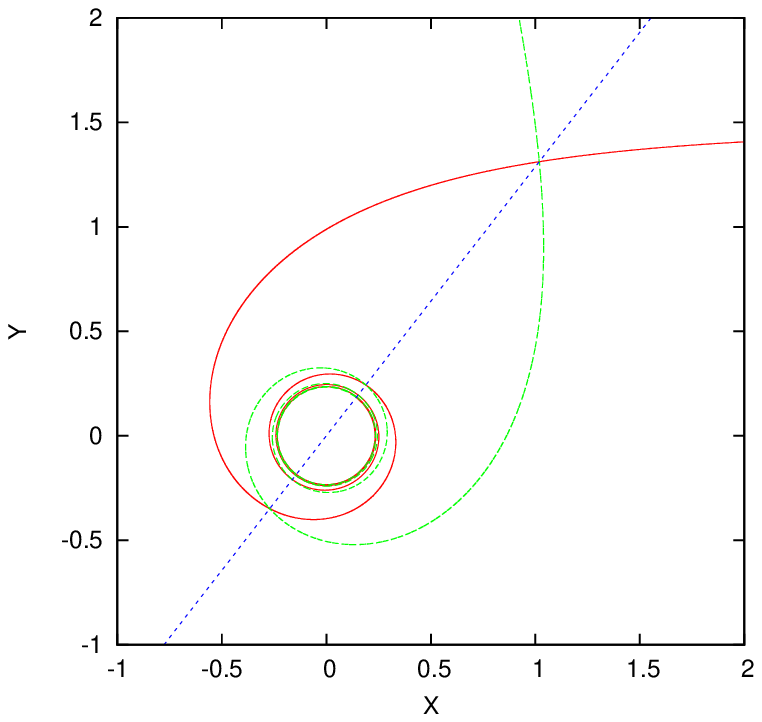}}
\caption{A trajectory in the center of mass frame for attractive interaction $\ga=2.05$ and $b/b_0=\beta+10^{-6}$ 
(only a portion of the trajectory is plotted). The first half part of the trajectory --- from $x=+\infty$ to the axis of symmetry  --- is 
plotted in red, the other half of the trajectory in green. 
The points of intersection of the trajectory lie on the axis of symmetry.}
\label{gamma2.05}
  \end{center}
\end{figure}

When $ \ga > 2 $, the angle $\phi$ diverges logarithmically for $ b $ approaching $\beta b_0 $. 
If we compare to the case ($ \ga = 2^+ $ and $ b/ b_0 \ll 1 $) previously studied, 
we see that the main difference is that now, the distance of closest approach $r_{min}$
is no longer small but of order one since $ r_{min} (b) \approx r_{min} (\beta b_0 ) 
= R = b_0 ( 2- \ga)^{1/ \ga } > 0 $ (see $\S$~\ref{gamma-lager-2}). 
The shape of the trajectories is therefore very different from the case 
($ \ga = 2^+ $ and $ b/ b_0 \ll 1 $) since then, the particle remains in a
(close to circular) orbit of positive radius. For the particular value $ b = \beta b_0$, the particle 
remains asymptotically trapped on a closed circular orbit, that is we have the formation of 
a binary. We illustrate this behavior in Fig.~\ref{gamma2.05}.

\section{Effect of a short-scale regularization in the potential}
\label{theo-soft}

In many physical situations, the potential is not a pure power-law as
in Eq.~\eqref{pot-def} but there is a regularization at small scales,
which is commonly called {\it{softening}} e.g. in the astrophysical
literature. This is for example the case in the dark--matter
collisionless N-body cosmological simulations, in which a softening is
introduced to minimize as much as possible collisional effects. From a
more fundamental point of view, we are interested in answering the
following questions:
\begin{enumerate}
\item Does a regularization in the potential modify the results above presented?

\item If yes, up to what scale and how?

\item Is the formation of pairs (which appears for $\ga>2$ and $b/b_0<\beta$) suppressed when 
a regularization is introduced, and in the affirmative case is there a minimal softening case needed?
\end{enumerate}

In this section we will answer these questions. In order to be able to make 
explicit calculations, we will consider two popular  regularization used commonly in the astrophysical literature (see e.g. \cite{Athanassoula_relax_2001,springel_05}), the {\it Plummer potential}
\be
\label{Plummer-potential}
v^{\text{Pl}}(r,\epsilon)=\frac{g}{\left(r^2+\ep^2\right)^{\gamma/2}} 
\ee
and the compact softening
\be
\label{compact-potential}
v^{\text{co}}(r,\epsilon) = \left\{ 
\begin{array}{ll}
\dfrac{g}{r^\gamma } & \text{if } r \geq \ep \\
 \dfrac{g}{\ep^\ga} \tv \left (r/\ep \right) & \text{if } 0 \leq r \leq \ep,
\end{array}
\right.
\ee
where $ \tv $ is a function on $ [ 0,1 ] $ such that $ \tv (1) = 1
$. The Plummer softening gives rise to an interaction of the same sign
than the unsoftened one, whereas it is possible to choose the
form of the compact softening (typically $\tv$ is a polynomial in
$r/\epsilon$) in order to be indifferently attractive, repulsive or
both.  A common feature for both of these potentials is that they
fulfill the relation
\be
 v (r,\epsilon) = \frac{g}{\ep^\ga} \mathcal{V} \left( \frac{r}{\ep} \right) ,
\ee
with
$$
 \mathcal{V}^{\text{Pl}} (R) = \frac{1}{ (R^2 + 1)^{\ga/2} }
$$
and
$$
 \mathcal{V}^{\text{co}} (R) = 
 \left\{ \begin{array}{ll}
 \dfrac{1}{R^\gamma } & \text{if } R \geq 1 \\
 \tv (R) & \text{if } 0 \leq R \leq 1 .
\end{array}
\right.
$$
We will show that the results presented below do not depend qualitatively on the explicit 
form of the regularization used. In what follow, we will study how the angle $\phi$ is modified by the regularization in the potential, first for repulsive interactions and then for attractive ones.

We introduce the angle $\phi_\epsilon$ corresponding to the regularized potential:
\be
\label{expressphisoft}
 \phi_\ep ( b , b_0 ) = \displaystyle
 \frac{b}{ r_{min} } \int_0^1 \frac{dx}{ \sqrt{1 - ( \frac{bx }{r_{min}} )^2 
  \pm \frac{2 b_0^\ga}{\ep^\ga} \mathcal{V}( \frac{r_{min} }{ \ep x} ) } } .
\ee

\subsection{Repulsive interactions with Plummer softening}
\label{sect-hard-coll-repulsive-Plumsoftening}

Here $ \mathcal{V}(R) = \mathcal{V}^{\text{Pl}} (R) = ( R^2 +1 )^{-\ga /2 } $. 
Then, the function $ r \mapsto 1 - b^2/r^2 - 2 b_0^\ga / \left( r^2 + \ep^2 \right)^{ \gamma/2} $ 
increases from $ - \infty $ to $ 1 $ as $r$ increases from $ 0^+ $ to $ +\infty $, hence 
has a single positive zero $ r_{min} $. It is easily checked that $ r_{min} $ 
is an increasing function of $b$ and that the function 
$ r \mapsto 1 - 2 b_0^\ga /\left( r^2 + \ep^2 \right)^{\gamma/2} $ possesses a positive zero 
if and only if $ \ep < b_0 2^{1 / \gamma } $. Therefore, for small $b$,
$$
 r_{min} \approx r_0 = b_0 2^{1 / \gamma } \sqrt{ 1 - \hat{\ep}^2 } , 
 \quad \quad \text{where} \ \hat{\ep} = \frac{\ep}{2^{1 / \gamma } b_0 } ,
$$
if $ \hat{\ep} \le 1 $, and
$$ 
 r_{min} \approx \frac{b}{\sqrt{1- \hat{\ep}^{-\ga} }} 
$$
if $ \hat{\ep} > 1 $. This naturally leads us to distinguish the case $ \ep < b_0 2^{1 / \gamma } $ 
and the case $ \ep > b_0 2^{1 / \gamma } $.

\subsubsection{The case $ \ep < b_0 2^{1 / \gamma } $}

We assume $ \hat{\ep} < 1 $, so that $ r_0 > 0 $, $ r_{min} = r_0 ( 1 + \mathcal{O}( (b/b_0)^2 ) ) $, 
and consider here again the small parameter $ \de = ( b/r_{min})^2 \ll 1 $. 
Substituting
$$ 
 \frac{2 b_0^\ga }{ \ep^\ga } = \frac{ 1 - b^2 / r_{min}^2 }{ \mathcal{V} ( r_{min} / \ep ) } 
 = \frac{ 1 - \de }{ \mathcal{V} ( r_{min} / \ep ) }
$$
yields
\begin{align*}
 & \phi_\ep ( b , b_0 )  = \displaystyle
 \sqrt{\de} \int_0^1 \frac{dx}{ \sqrt{1 - \de x^2 
  - ( 1 - \de )\frac{ \mathcal{V} ( r_{min} / ( \ep x) ) }{ \mathcal{V} ( r_{min} / \ep ) } } } 
 \\ 
 & = \sqrt{\de} \int_0^1 \frac{dx}{ \sqrt{ F(x,r_{min} /\ep ) + \de ( 1 - x^2 - F(x, r_{min} /\ep ) ) } } ,
\end{align*}
where we have set
$$
 F (x,r_{min} /\ep ) = 1 - \frac{ \mathcal{V} ( r_{min} / ( \ep x) ) }{ \mathcal{V} ( r_{min} / \ep ) } .
$$
We prove in App.~\ref{appbornage1} that the function $ x \mapsto \frac{1-x^2}{F(x, r_{min} /\ep ) } $ 
is bounded on $[0,1] $ independently of $b$. This shows that we may apply the argument for 
Eq.~\eqref{prototype} and write
\begin{align*}
 \phi_\ep ( b , b_0 ) & = \sqrt{\de} \int_0^1 \frac{dx}{ 
  \sqrt{ F (x,r_{min} / \ep) } \sqrt{ 1 + \de ( \frac{ 1 - x^2 }{ F(x, r_{min}/\ep)} - 1 ) } }
 \\ 
 & = \sqrt{\de} \int_0^1 \frac{dx}{ \sqrt{ F(x,r_{min} /\ep ) }}  + \mathcal{O} ( \de^{3/2} ) .
\end{align*}
At this stage, since $ r_{min} = r_0 ( 1 + \mathcal{O}( (b/b_0)^2 ) ) $, 
one could legitimate the expansion
$$
 \int_0^1 \frac{dx}{ \sqrt{ F(x, r_{min}/\ep) } } 
 = \int_0^1 \frac{dx}{ \sqrt{ F(x, r_0/\ep) } } + \mathcal{O} ( (b/b_0)^2 ).
$$
Since $ r_{min} = r_0 ( 1 + \mathcal{O}( (b/b_0)^2 ) ) $, 
$ \sqrt{\de} = b / r_{min} = b /r_0 ( 1 + \mathcal{O}( (b/b_0)^2 ) ) $, 
and thus, when $ \hat{\ep} < 1 $,
\be
\label{hardrepulssoft}
\phi_\ep^{\text{Pl}} ( b , b_0 ) = B^{\text{Pl}}_{\hat{\ep}} (\gamma) (b / b_0 ) 
+ \mathcal{O} ( (b/b_0)^3 ) ,
\ee
where
$$
 B^{\text{Pl}}_{\hat{\ep}} ( \gamma ) = \frac{ 2^{ -1/\ga} }{ \sqrt{ 1 - \hat{\ep}^2 } } 
 \int_0^1 \frac{dx}{ \sqrt{ 1 - \dfrac{x^\ga}{ ( 1 - \hat{\ep}^2 ( 1- x^2) )^{\ga/2} } } } .
$$
Comparing Eq.~\eqref{hardrepulssoft} with the expression Eq.~\eqref{durmoins} of 
the angle of closest approach without softening we observe that the linear dependence 
of $\phi$ with respect to $b / b_0$ is not modified, only the pre-factor changes. 
It is also easy to check that in the limit $\epsilon\to0$ we have, as expected, 
$ B^{\text{Pl}}_{\hat \ep} (\gamma ) \to B ( \gamma ) $. 
As expected, the new introduced scale is $\epsilon$.

\subsubsection{The case $ \ep > b_0 2^{1 / \gamma } $}
\label{grandepsi}

In the case $ \hat \ep > 1 $, that is $ \ep > 2^{ 1/\ga} b_0 $, we recall that 
\be
\label{develo44}
 r_{min} \approx b / \sqrt{ 1 - \hat{\ep}^{-\ga} } 
\ee
and that
$$
 \phi_\ep ( b , b_0 ) = \displaystyle
 \frac{b}{ r_{min} } \int_0^1 \frac{dx}{ \sqrt{1 - ( \frac{bx }{r_{min}} )^2 
  - 2 \frac{ b_0^\ga}{\ep^\ga} \mathcal{V}( \frac{r_{min} }{ \ep x} )} } .
$$
Substituting 
$ 1 = b^2 / r_{min}^2 + \hat{\ep}^{-\ga} \mathcal{V}( r_{min}/ \ep) $ in 
the integral and considering the small parameter $ \de = r_{min}^2 / \ep^2 \sim b^2 / \ep^2 $ gives
$$
 \phi_\ep ( b , b_0 ) = \int_0^1 \frac{dx}{ \sqrt{ G_b (x) } } ,
$$
where
\begin{align*}
 G_b (x) = 1 - x^2 
- \frac{r_{min}^2 }{b^2 \hat{\ep}^\ga} 
 \left( \mathcal{V}( \sqrt{\de}/x ) - \mathcal{V}( \sqrt{\de} ) \right) .
\end{align*}
In view of the fact that $ r_{min}^2 \approx b^2 / ( 1 + \hat{\ep}^{-\ga} ) $ 
and $ b_0 \lesssim \ep $, we expect
$$
 \phi_\ep ( b , b_0 ) \approx \int_0^1 \frac{dx}{ \sqrt{1 - x^2 } } = \frac{\pi}{2} .
$$
We also see that the situation is similar to the form given in Eq.~\eqref{prototype}, but the dependency 
on the small parameter $ \de $ is more intricate. Actually, for the Plummer potential, we have 
$ \mathcal{V}^{\text{Pl}} (R) = ( R^2 + 1 )^{-\ga /2 } $, thus, for small $R$, 
$ \mathcal{V}^{\text{Pl}} (R) = 1 - \ga R^2 /2 + \mathcal{O}(R^4 ) $. Therefore, 
for fixed $x$ and small $\de $, we obtain
$$
G_b (x) = 1 - x^2 - \frac{ \ga \de }{ 2( \hat{\ep}^{\ga} - 1) } 
\left( \frac{1}{x^2} - 1 \right) + \mathcal{O} ( \de^2 ) ,
$$
which is a situation very similar to the form given in Eq.~\eqref{prototype}, 
but unfortunately the function $ x \mapsto ( 1/x^2 - 1)/( 1 - x^2 ) = - 1 / x^2 $ 
being too singular near the origin, the power series expansion trick used for Eq.~\eqref{prototype} 
as in Subsect.~\ref{section-softcoll} and \ref{sect-hard-coll-repulsive} breaks down. 

Following the approach used in $\S$~\ref{gammadeuxtiersdeux}, we 
divide the correction $ \phi_\ep ( b , b_0 ) - \pi /2 $ by $\de$ and write it under the form
\begin{align*}
 - \frac{1}{ \de } \left( \phi_\ep ( b , b_0 ) - \frac{\pi}{2} \right)
 & 
 = 2 \frac{r_{min}^2 b_0^\ga \ep^2 }{b^2 } \int_0^1 g(x) \, dx 
 \\ 
 & \approx \frac{1 }{ \hat{\ep}^{\ga} - 1 } \int_0^1 g(x) \, dx
\end{align*}
by Eq.~\eqref{develo44} and with
$$
 g(x) = \frac{ \mathcal{V}( \sqrt{\de} ) - \mathcal{V}( \sqrt{\de}/x ) }{ 
 \de \sqrt{G_b(x)} \sqrt{1-x^2} [ \sqrt{G_b(x)} + \sqrt{1-x^2} ] } \ge 0 .
$$
Clearly, as $ b / \ep $ goes to $0$, $ \de \ll 1 $, $ G_b (x) \approx 1 - x^2 $ and we have
$$
 \int_0^1 g (x) \, dx \to \frac{ \ga }{4} 
 \int_0^1 \frac{ \frac{1}{x^2} - 1 }{ (1-x^2)^{3/2} } \, dx = +\infty ,
$$
due to the non integrable singularity at the origin. We shall prove that actually 
$ \int_0^1 g(x) \, dx \sim \de^{-1/2} $. As a first step, as in $\S$~\ref{gammadeuxtiersdeux}, 
we get rid of the contribution for $ 1 /2 \leq x \leq 1 $. 
Indeed, $ \int_0^1 g (x) \, dx \to +\infty $ whereas
$$ 
 \int_{1/2}^1 g (x) \, dx \to 
 \frac{ \ga }{4} \int_{1/2}^1 \frac{ \frac{1}{x^2} - 1 }{ (1-x^2)^{3/2} } \, dx < +\infty .
$$
As a consequence, using the natural substitution $ y = \sqrt{\de} / x $, 
\begin{align*}
\int_0^1 g(x) \, dx & \approx \int_0^{1/2} g (x) \, dx 
\\
& = \frac{1}{ \sqrt{\de} } \int_{ 2 \sqrt{\de} }^{+\infty} 
\frac{ \mathcal{V}( \sqrt{\de} ) - \mathcal{V}( y ) }{ D_b(y) } \, dy
\end{align*}
where we have denoted
\begin{align*}
 D_b (y) & = y^2 \sqrt{ G_b\left( \frac{\sqrt{\de} }{ y} \right) \left(1- \frac{\de}{y^2 } \right) }  
 \\ & \quad \times 
 \left[ \sqrt{G_b\left( \frac{\sqrt{\de} }{ y } \right)} + \sqrt{ 1 -\frac{\de}{ y^2 }} \right] . 
\end{align*}
When $ \de \to 0 $, we have
$$
 G_b( \sqrt{\de} / y ) \to G_{\hat{\ep}}^- (y) 
 = 1 - \frac{ 1 }{ \hat{\ep}^{\ga} -1} \left( \mathcal{V}(y) - \mathcal{V}(0) \right)
$$
and one could rigorously justify that
\begin{align*}
& \int_0^1 g(x) \, dx \approx 
\frac{1}{ \sqrt{\de} } \int_0^{+\infty} \frac{  \mathcal{V}( 0 ) -  \mathcal{V}(y ) }{ 
y^2 \sqrt{G_{\hat{\ep}}^- (y)} [ \sqrt{G_{\hat{\ep}}^-(y)} + 1 ] } \, dy 
\\
& 
= \frac{ \hat{\ep}^{\ga} -1}{ \sqrt{\de} } \int_0^{+\infty} 
\left( 1 - \frac{1}{\sqrt{ 1 + \frac{1}{ \hat{\ep}^{\ga} -1} \left( \mathcal{V}(0) - \mathcal{V}(y) \right)} } \right) \, \frac{dy}{y^2} .
\end{align*}
The last integral is indeed convergent since: for large $ y $, $\mathcal{V} (y) \to 0 $, 
thus the integrand is $ \sim 1 / y^2 $; for small $y$, 
$  \mathcal{V}^{\text{Pl}}(0) - \mathcal{V}^{\text{Pl}}(y ) = 
1 - (1 +y^2 )^{- \ga/2} \approx \ga/ ( 2y^2) $, thus the integrand is continuous 
at the origin. It then follows that, for $ b \ll \ep $:
\begin{align}
\label{hardrepulssoft-autre}
\phi_\ep ( b , b_0 ) & 
= \frac{\pi}{2} 
- \tilde{B}_{\hat{\ep}}^{\text{Pl}} (\gamma ) b /\ep + o( b / \ep ) ,
\end{align}
with
\begin{align*}
 & \tilde{B}_{\hat{\ep}}^{\text{Pl}} ( \gamma ) 
 = \frac{ 1 }{ \sqrt{ 1- \hat{\ep}^{-\ga} } } \\
 & \quad \times \int_0^{+\infty} 
\left( 1 - \frac{1}{\sqrt{ 1 + \frac{1}{ \hat{\ep}^{\ga} -1} 
\left( \mathcal{V}^{\text{Pl}}(0) - \mathcal{V}^{\text{Pl}}(y) \right)} } \right) \, \frac{dy}{y^2} 
 > 0 .
\end{align*}
If $ \ep \gg b_0 $, that is $ \hat{\ep} \gg 1 $, we justify in Appendix~\ref{app14} that
\be
\label{integrale14}
 \tilde{B}_{\hat{\ep}}^{\text{Pl}} ( \gamma ) \approx 
 \hat{\ep}^{-\ga} \sqrt{\pi} \frac{ \Gamma \left( \frac{\ga+1}{2} \right) }{ 
 4\Gamma \left( \frac{\ga}{2} \right)} .
\ee

We see here that, because $\epsilon\gg b_0$, the value of $\phi$ is completely different compared 
to the case $\epsilon\to0$. As expected,  in the limit $b\to0$, $\phi\to\pi/2$, which means that 
the particle trajectory is unperturbed.

\subsection{Repulsive interactions with compact softening}
\label{sect-hard-coll-repulsive-compsoftening}

In this Subsection, we give the few modifications appearing in the asymptotic expansions 
when we consider a compact softening Eq.~\eqref{compact-potential}. The formula 
we shall obtain are qualitatively comparable to those in 
Subsect.~\ref{sect-hard-coll-repulsive-Plumsoftening} for the Plummer softening. 
The first step is to determine the asymptotic behavior of $ r_{min} $, and here again, 
we shall distinguish the cases where $ \hat{\ep} = \ep / ( 2^{1/\ga} b_0 ) $ is small or large.

\subsubsection{The case $ \ep < b_0 2^{1 / \gamma } $} 

Assume that $ \ep < b_0 2^{1 / \gamma } $, that is $ \hat{\ep} = \ep /(b_0 2^{1/\gamma} ) < 1 $. 
Then, the function $ r \mapsto 1 - b^2/r^2 - 2 b_0^\ga / r^\ga $ is 
increasing on $ [ \ep , +\infty ) $. It follows that this function has a unique zero $r_{min} $ 
on $ [ \ep , +\infty ) $, which satisfies, for $ b / b_0 \ll 1 $,
$$
 r_{min} \approx b_0 2^{1/\ga} > \ep .
$$
In view of the fact that $ r_{min} \approx b_0 2^{1/\ga} > \ep $, the trajectory never 
enters in the region $\{ r \leq \ep \} $ where the softening has an effect, hence we 
obtain the same asymptotics as in the case without softening (see Eq.~\eqref{durmoins}), 
namely
\be
\label{hardrepulssoftcompact}
\phi_\ep ( b , b_0 ) = B (\gamma) (b / b_0 ) + \mathcal{O} ( (b/b_0)^3 ) ,
\ee
where $ B (\gamma) $ is the same as in Eq.~\eqref{durmoins}.

\subsubsection{The case $ \ep > b_0 2^{1 / \gamma } (\max_{\mathbb{R}} \mathcal{V} )^{1/\ga} $}

Assume now that $ \ep > b_0 2^{1 / \gamma } ( \max_{\mathbb{R}} \mathcal{V} )^{1/\ga} $, 
that is $ \hat{\ep}^\ga = \ep^\ga /(2 b_0^{\gamma}) > \max_{\mathbb{R}} \mathcal{V} 
= \max_{[0,1]} \mathcal{V} \geq 1$. 
The function $ r \mapsto 1 - b^2/r^2 - 2 b_0^\ga/r^\ga $ is then increasing on 
$ [ \ep , +\infty ) $ from $ 1 - b^2/\ep^2 - \hat{\ep}^{-\ga} $ to $ 1$. 
Since $ \hat{\ep} > 1 $, we have, for $ b \ll \ep $, 
$ 1-b^2/\ep^2 - \hat{\ep}^{-\ga} \approx 1 - \hat{\ep}^{-\ga} > 0 $, 
hence $ 1 - b^2/r^2 - 2 b_0^\ga/r^\ga $ is positive on $ [\ep ,+ \infty ) $. 
On $ [0, \ep ] $, the function 
$ r \mapsto 1 - b^2/r^2 - \hat{\ep}^{-\ga} \mathcal{V} (r/\ep) $ is $> 0 $ 
for $r = \ep $ and tends to $ - \infty $ for $r \to 0 $, thus has a largest root $r_{min} \leq \ep$. 
Moreover, since $ b^2 / r_{min}^2 = 1 - \mathcal{V} (r_{min}/\ep) \hat{\ep}^{-\ga} 
\geq 1 - \hat{\ep}^{-\ga} \max_{\mathbb{R}} \mathcal{V} > 0 $ by our hypothesis, we have 
$ r_{min} \lesssim b \ll \ep $, hence 
$$ 
 r_{min} 
 = \frac{b }{\sqrt{ 1 - \mathcal{V} (r_{min} /\ep) \hat{\ep}^{-\ga} } } 
 \approx \frac{b }{\sqrt{ 1 - \mathcal{V} (0) \hat{\ep}^{-\ga} } } 
$$
that is close to Eq.~\eqref{develo44}. We may then carry out computations very similar to 
those leading to Eq.~\eqref{hardrepulssoft-autre}, provided $ \tv $ is $ \mathcal{C}^2 $ 
on $[0,1 ] $, positive on $ ( 0, 1 ] $ and $ \tv'(0) = 0 $. This yields
\begin{align}
\label{hardrepulssoftco-autre}
\phi_\ep ( b , b_0 ) & 
= \frac{\pi}{2} 
- \tilde{B}_{\hat{\ep}}^{\text{co}} (\gamma ) b /\ep + o( b / \ep ) ,
\end{align}
with
\begin{align*}
 & \tilde{B}_{\hat{\ep}}^{\text{co}} ( \gamma ) 
 = \frac{ 1 }{ \sqrt{ 1- \mathcal{V}^{\text{co}}(0) \hat{\ep}^{-\ga} } } \\
 & \quad \times \int_0^{+\infty} 
\left( 1 - \frac{1}{\sqrt{ 1 + \frac{1}{ \hat{\ep}^{\ga} -\mathcal{V}^{\text{co}}(0)} 
\left( \mathcal{V}^{\text{co}}(0) - \mathcal{V}^{\text{co}}(y) \right)} } \right) \, \frac{dy}{y^2}.
\end{align*}
Here, we do not claim that $ \tilde{B}_{\hat{\ep}}^{\text{co}} ( \gamma ) $ is a 
positive constant. For instance, if $ \tv(0) = 0 $, then $ \tilde{B}_{\hat{\ep}}^{\text{co}} ( \gamma ) < 0 $, 
whereas if $ \tv(x) = \tv(0) $ on $ [ 0, 1 ] $, then $ \tilde{B}_{\hat{\ep}}^{\text{co}} ( \gamma ) > 0 $. 
For a general function $ \tv $ on $ [ 0,1 ] $, it may happen exceptionally that 
$ \tilde{B}_{\hat{\ep}}^{\text{co}} ( \gamma ) $ vanishes, and in this case, the correction 
$ \phi_\ep - \pi /2 $ is not of order $ b /\ep $ but smaller. This however does 
not happen for generic functions $ \tv $.

\subsection{Attractive interactions with a softening}
\label{sect-hard-coll-attractive-softening}

The function $ r \mapsto 1- b^2 / r^2 + \hat{\ep}^{-\ga} \mathcal{V} \left( r/ \ep \right) $ 
tends to $ 1$ at infinity and to $ - \infty $ at $ 0^+ $, hence possesses a larger zero $ r_{min} $, 
but there may exist several zeros in general. We shall prove that independently whether 
$ \ep / b_0 $ is small or not, we have
\be
\label{develo45}
 r_{min} \approx \frac{b}{ \sqrt{1 + \mathcal{V}(0) \hat{\ep}^{-\ga} } } 
\ee
(whereas without softening, we had $ r_{min} \sim b^{2/(2-\ga)} $), and
\be
\label{attractifhard-soft-autre}
\phi_\ep ( b , b_0 ) 
= \frac{\pi}{2} + C_{\hat{\ep}} (\ga ) \frac{b}{ \ep } + o( b / \ep ) ,
\ee
where
\begin{align*}
 & C_{\hat{\ep}} (\ga ) 
 = \frac{ 1 }{ \sqrt{ 1 + \mathcal{V}(0)\hat{\ep}^{-\ga} } } 
 \\
& \quad \times \int_0^{+\infty} \left( \frac{1}{ \sqrt{1 - \frac{1}{ \hat{\ep}^\ga +\mathcal{V}(0) } (
\mathcal{V}(0) - \mathcal{V}(y) ) } } - 1 \right) \, \frac{dy}{y^2} .
\end{align*}
If $ \ep \gg b_0 $, that is $ \hat{\ep} \gg 1 $, we can show (as we have done for 
Eq.~\eqref{integrale14}) that
\be
\label{C-large-softening}
C_{\hat{\ep}} (\ga ) \approx \hat{\ep}^{-\ga} \sqrt{\pi} 
\frac{ \Gamma \left( \frac{ \ga +1 }{2} \right) }{ 4 \Gamma \left( \frac{\ga}{2} \right)}.
\ee
On the other hand, if $ \ga < 2 $ and $ \ep \ll b_0 $, that is $ \hat{\ep} \ll 1 $, we 
can show that
$$
 C_{\hat{\ep}} (\ga ) \approx \frac{\hat{\ep}^{\ga/2}}{\sqrt{\mathcal{V}(0)}} 
 \int_0^{+\infty} \left( \sqrt{\frac{\mathcal{V}(0)}{ \mathcal{V}(y)} } - 1 \right) \, \frac{dy}{y^2} .
$$
We have then a big difference with the case of repulsive interactions studied in 
Sect.~\ref{sect-hard-coll-repulsive-compsoftening} (and also 
in Sect.~\ref{sect-hard-coll-repulsive-Plumsoftening}), where 
$ \phi_\ep \sim b / \max ( \ep, 2^{1/\ga} b_0 ) $, displaying the characteristic length 
$ \ep $ or $ b_0$ depending which one is the largest one. Here, for attractive interactions, 
only the softening characteristic length $ \ep $ appears in the first order term 
$ \phi_\ep - \pi / 2 \sim b / \ep $ in Eq.~\eqref{attractifhard-soft-autre}. This point 
is not completely surprising in view of the different cases appearing in 
Subsect.~\ref{sect-hard-coll-attractive} (for $ \ga < 2/3 $, $ \ga = 2/3 $, $ 2/3 < \ga < 2 $, etc), 
which correspond very roughly to the case $ \ep = 0 $.

Since $ 1 \le 1 + \hat{\ep}^{-\ga} \mathcal{V} \left( r_{min}/ \ep \right) = b^2/ r_{min}^2 $, 
we must have $ r_{min} \le b \ll \ep $, and this in turn implies 
Eq.~\eqref{develo45}.

Our small parameter here will be $ \de = r_{min}^2 / \ep^2 \ll 1 $ (by Eq.~\eqref{develo45}). 
Substituting 
$ 1 = b^2 / r_{min}^2 - \hat{\ep}^{-\ga} \mathcal{V}( r_{min}/ \ep ) $ in the integral gives
$$
 \phi_\ep ( b , b_0 ) =  \int_0^1 \frac{dx}{ \sqrt{ G_b (x) } } ,
$$
where
$$
 G_b (x) = 1 - x^2 
 + \frac{r_{min}^2 }{b^2 \hat{\ep}^\ga} 
 \left( \mathcal{V}( \sqrt{\de}/x ) - \mathcal{V}( \sqrt{\de} ) \right) .
$$
Comparing with $\S$ \ref{grandepsi}, the only difference is a change of sign. 
Therefore, similar computations to those in that paragraph yield Eq.~\eqref{attractifhard-soft-autre}.

\subsection{Computation of a threshold in $\ep $ for attractive potentials with $ \ga > 2 $}

When $ \ga > 2 $ and without softening in the potential (formally, $\ep = 0 $), 
the deflection angle $ \phi $ diverges logarithmically to $ + \infty $ when 
$ b > \beta b_0 $ approaches $ \beta b_0 $ (see Eq.~\eqref{phi-g-gt2}). 
This divergence is due to the fact that $ r_* \approx R = b_0 ( 2 - \ga)^{1/\ga} $ 
becomes a double root of the function $ W $ in this limit. The first paragraph of this 
Subsection is devoted to the proof of the existence of some threshold 
$ \ep_* ( b_0 , \ga ) > 0 $, for the Plummer softening, such that 
if $ \ep < \ep_*( b_0, \ga ) $, then the angle $ \phi_\ep $ still diverges for 
some specific value of $r$ (depending on $ b_0 $, $\ga $ and $\ep $), whereas 
for $ \ep > \ep_*( b_0, \ga ) $, the angle $ \phi_\ep $ no longer diverges and 
is a smooth function of $ b / b_0 $ for all positive values of $ b / b_0 $. 
This means that in order to remove the divergence in $ \phi $, one has to use 
a sufficiently large softening. 
In the first case, the divergence is here again due to the existence of some 
positive double root in $r$ for the function
$$
 W_{b,\ep} (r) = 1 - \frac{b^2}{r^2} + \hat{\ep}^{- \ga} \mathcal{V} \left( \frac{r}{\ep} \right) ,
$$
whereas for $ \ep > \ep_*( b_0, \ga ) $, the function $ W_{b,\ep} (r) $ has no double root. In the second paragraph we will discuss the case of the compact softening.

\subsubsection{The case of a Plummer softening}

We now consider the Plummer softening $\mathcal{V}(R) = \mathcal{V}^{\text{Pl}} (R) 
= ( 1 +R^2)^{-\ga/2} $ and are interested in determining under which condition on $\ep$ the 
function $ W_{b,\ep} $ has a unique zero $ r_{min} $ for any $b>0$. We have
$$
 W'_{b,\ep} (r) = \frac{2 \ga b_0^\ga}{r^3} \left( 
 \frac{b^2}{\ga b_0^\ga} - \frac{ r^4}{ ( r^2 + \ep^2 )^{\ga/2 +1 } } \right) 
$$
and, denoting $ r = \ep R $,
$$
 \frac{ r^4}{ ( r^2 + \ep^2 )^{\ga/2 +1 } } = \ep^{ 2 - \ga } \frac{ R^4}{ ( R^2 + 1 )^{\ga/2 +1 } } .
$$
The function $ R \mapsto R^4/ ( R^2 + 1 )^{\ga/2 +1 } $ is increasing 
on $ [ 0, R_{max} ] $ and decreasing on $ [ R_{max} , +\infty ) $ (recall $ \ga > 2 $), 
where $ R_{max} = \sqrt{4 / ( \ga - 2 ) } $; its maximal value is 
$ M (\ga ) = 16 (\ga-2)^{\frac{\ga}{2}-1} (\ga+2)^{-\frac{\ga}{2}-1} $. 
Therefore, when $ b^2 / ( \ga b_0^\ga ) < \ep^{ 2 - \ga } M (\ga ) $ (case 1), the 
function $ W_{b,\ep} $ is increasing on $ ( 0, r_1 ] $, decreasing on $ [ r_1 , r_2 ] $ 
and increasing on $ [ r_2 , + \infty ) $ 
when $ b^2 / (\ga b_0^\ga) > \ep^{ 2 - \ga } M (\ga ) $ (case 2), the function 
$ W_{b,\ep} $ is increasing on $ ( 0, + \infty ) $.
The two critical points $ r_1$ and $r_2$ merge for $ b^2 / (\ga b_0^\ga) = \ep^{2-\ga} M (\ga) $, 
and we shall see that the threshold is determined by the sign of $ W_{b,\epsilon} $ at this merging 
point $r_1 = r_2 $.

Let us now fix $ \ep > 0 $. For $b $ very small, we are in case 1 and
the two positive roots $r_1 $ and $r_2 $ of the equation $ b^2/(\ga
b_0^\ga) = r^4 / ( r^2 + \ep^2 )^{\ga/2 +1 } $ are $ r_1 $ very small
and $ r_2 $ very large. The function $ W_{b,\ep} $ has then a local
minimum $ W_{b,\ep} (r_2) \approx 1 $. When $b$ increase, $ W_{b,\ep}
$ decrease, the two critical points $ r_1$ and $r_2$ merge when $
b^2/(\ga b_0^\ga) = \ep^{ 2 - \ga } M (\ga ) $, and for larger $b$,
$W_{b,\ep} $ is increasing on $ ( 0,+\infty ) $.

Let us consider the special value of $b_{crit}$ where 
$ b_{crit}^2 / ( \ga b_0^\ga ) = \ep^{ 2 - \ga } M (\ga ) $, 
for which the two critical points $ r_1$ and $r_2$ merge: $ r_1 = r_2 = r_{crit} = \ep R_{max} $. 
If $ W_{b_{crit},\ep} ( r_{crit} ) > 0 $, then by monotonicity in $b$, 
for any $ b > 0 $, the function $ W_{b_{crit},\ep} $ has a single positive zero $ r_{min} $. 
If now $ W_{b_{crit},\ep} ( r_{crit} ) < 0 $, then, still by monotonicity in $b$, for 
$ b $ smaller, but close to $ b_{crit} $, $ W_{b, \ep } $ has two critical points 
$ 0 < r_1 < r_2 $ with $ 0 > W_{b, \ep } (r_1) > W_{b, \ep } (r_2) $. As $b$ decreases, 
the critical value $ W_{b, \ep } (r_2) $ will be zero for some particular value of $b = b_\sharp $ 
for which $r_2 $ has become a double root of $ W_{b_\sharp , \ep } $, yielding a logarithmic 
divergence in $ \phi_\ep $. As a consequence, we simply need to determine the sign of
\begin{align*}
 W_{b_{crit},\ep} ( r_{crit} ) 
 & = 1 - \frac{b_{crit}^2}{\ep^2 R_{max}^2 } 
 + \frac{2 b_0^\ga}{ ( \ep^2 R_{max}^2 + \ep^2 )^{\ga/2} } 
 \\
 & = 1 - \frac{\ep^{ - \ga } M (\ga ) \ga b_0^\ga }{ R_{max}^2 } 
 + \frac{2 b_0^\ga \ep^{-\ga} }{ ( R_{max}^2 + 1 )^{\ga/2} } 
 \\
 & = 1 - ( \ep_*(b_0, \ga ) / \ep )^\ga ,
\end{align*}
where the threshold is given by
\be
\label{seuil}
 \ep_*(b_0, \ga) = 2^{1/\ga} b_0 \left( \frac{\ga -2}{ \ga + 2 } \right)^{ \frac12 + \frac{1}{\ga} } .
\ee
It follows that if $ \ep > \ep_*(b_0, \ga) $, then $ \phi_\ep $ is a smooth function of $b$ 
see Fig.~\ref{gamma2.5eps}, whereas if $ \ep < \ep_*(b_0, \ga) $, then $ \phi_\ep $ diverges as 
$b$ approaches some value $ b_\sharp = b_\sharp (\ep ) $ corresponding to the case where 
$ W_{b, \ep } $ has zero as a local minimum. By computations very similar to those 
in Sect.~\ref{sect-hard-coll-attractive}, we see that the divergence is logarithmic. 
One may also check that if $ \ep = \ep_*(b_0, \ga) $, then $ \phi_\ep $ is a diverging function of $b$ for some $ b_\sharp = b_\sharp (\ep ) $.
In other words, in order to regularize the divergence in the case $ \ga  > 2 $, we have 
to use a sufficiently large softening parameter, namely $ \ep > \ep_*(b_0, \ga) $.

Let us finally consider the case $ \ga = 2 $. Notice that formally, 
$ \ep_* ( b_0 , \ga ) \to 0 $ as  $\ga \to 2 $, hence we may think that $ \phi_\ep $ is 
a smooth function of $b$ for any $ \ep > 0 $, and this is indeed the case. 
Actually, in the case $ \ga = 2 $, the function $ R \mapsto R^4 / ( R^2+1)^2 $ 
is increasing on $ [ 0, +\infty ) $, and tends 
to $1$ at infinity. Therefore, either $ b^2 / 2 b_0^2 < 1 $ and then the function 
$ W_{b,\ep} $ is increasing on $ ( 0, r_1 ] $ and decreasing on $ [ r_1 , +\infty )$; 
either $ b^2 / (2 b_0^2) \ge 1 $ and then the function $ W_{b,\ep} $ is increasing on 
$ ( 0, +\infty )$. In any case $ W_{b, \ep } $ has a single zero $ r_{min} $ and we never have 
a double root. It follows that $ \phi_\ep $ is a smooth function of $b$.

\subsubsection{The case of a compact softening}

For a general compact softening $ \mathcal{V} = \mathcal{V}^{\text{co}} $, 
computations are much less explicit. We first have
$$
 W_{b, \ep }'( r = \ep R ) 
 = \frac{2 b_0^\ga }{R^3 \ep^{\ga +1} } \left( \frac{b^2 \ep^{\ga-2} }{ b_0^\ga } 
 + R^3 \mathcal{V}'(R) \right) ,
$$
and we then need to know the behavior of the function $ R \mapsto - R^3 \mathcal{V}'(R) $, 
which certainly has a positive maximum $ M = M ( \tv ) $ attained at some $ 0 < R_{max} \le 1 $ 
since $ \ga >2 $. 
If the function $ R \mapsto - R^3 \mathcal{V}'(R) $ is increasing on $[ 0, R_{max} ] $ 
and then decreasing on $ [R_{max}, +\infty ) $, the behavior is the same as the one 
previously described for the Plummer softening. Since
\begin{align*}
 W_{b_{crit}, \ep} ( r_{crit} )
 & = 1 - \frac{b_0^\ga M(\tv) }{\ep^\ga R_{max}^2 } 
 + \frac{2 b_0^\ga}{\ep^\ga} \mathcal{V} ( R_{max} ) 
 \\
 & = 1 - \frac{b_0^\ga}{\ep^\ga } \left( \frac{M(\tv)}{R_{max}^2} - 2 \mathcal{V} (R_{max} ) \right) 
  \\
 & =
  1 + \frac{b_0^\ga}{\ep^\ga } \left( R_{max}\mathcal{V}' (R_{max} ) + 2 \mathcal{V} (R_{max} ) \right) ,
\end{align*}
there exists a threshold if and only if 
$ M(\tv)/ R_{max}^2 = - R_{max} \mathcal{V}' (R_{max} ) > 2 \mathcal{V} (R_{max} ) $, 
and otherwise, we never have a double root for $W_{b,\ep} $ hence no divergence 
in $ \phi_\ep $. The example below illustrate the first case.

If $ \ga = 3 $ and $ \tv(R) = 21 R^2 - 35 R^3 + 15 R^4$ for $ 0 \le R \le 1 $, then 
$ R \mapsto - R^3 \mathcal{V}'(R) $ is decreasing and negative on $ [ 0 , \approx 0.474] $, 
increasing on $ [ \approx 0.474 , \approx  0.984 ] $ and decreasing on 
$ [ \approx  0.984 , + \infty ) $, hence has maximum value $ M (\tv) \approx 3.023 $ attained 
at $ R_{max} \approx 0. 984 $. Moreover, 
$ M(\tv)/ R_{max}^2 - 2 \mathcal{V} (R_{max} ) \approx 1.023 > 0 $, thus the variations of 
$W_{b,\ep} $ are the same as for the Plummer softening, with a threshold given by 
$$ 
 \ep_* (b_0 , \ga ) = b_0 \left(\frac{M(\tv)}{R_{max}^2} - 2 \mathcal{V} (R_{max} ) \right)^{1/3} 
 \approx 1.0077 b_0.
$$

\subsection{Summary of the results and numerical checking}
\label{summary-soft}
We  summarize in table \ref{yellowsummary-softening} the results obtained in this section. We have shown that the effect of the softening does not depend strongly on the form of the softening, obtaining the same qualitative results for the two softening considered --- Plummer and compact one. There is an exception for repulsive interactions and $\epsilon < b_0 2^{1/\ga}$, in which case the compact softening does not modify the trajectory of the particles because they do not visit the region in which the potential is regularized.

\begin{table*}
\begin{tabular}{| c || c |}
 \hline
 repulsive potential & attractive potential \\ 
 \hline \hline 
 \begin{tabular}{c | c }
 $ \phi_\ep  \sim b / b_0 $ \quad  when $ b \ll b_0 $ & 
 if \quad $ \displaystyle \hat{\ep} = \ep / ( 2^{1/\ga} b_0 ) < 1 $ 
 \\ \hline 
 $ \phi_\ep - \pi /2 \sim -b / \ep  $ \quad when $ b \ll \ep $ & 
 if \quad $ \hat{\ep} = \ep / ( 2^{1/\ga} b_0 ) > 1$
 \end{tabular} 
 &  $ \phi_\ep - \pi /2 \sim b / \ep  $  \quad when $ b \ll \ep $ 
 \\ \hline 
\end{tabular}
\caption{Summary of the expansions of the angle $\phi_{\ep} $ with a Plummer 
softening in the potential for hard collisions}
\label{yellowsummary-softening}
\end{table*}

\begin{figure}
  \begin{center}
    \psfrag{X}{$\epsilon/b_0$}
    \psfrag{Y}[c]{$\phi_\ep(\epsilon/b_0)$}
    	\psfrag{AAAAAAAA}[c][c]{$\gamma=1/2$}
  	\psfrag{BBBBBBBB}[c][c]{$\gamma=1$}
  	\psfrag{CCCCCCCC}[c][c]{$\gamma=3/2$}
   	{\includegraphics[height= 0.35\textwidth]{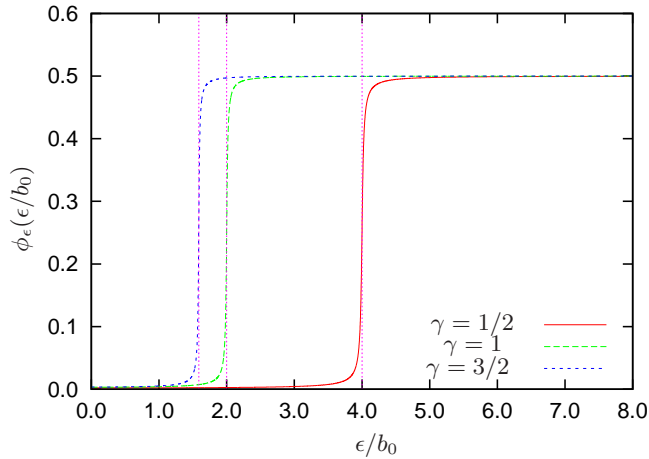}}\\
    \caption{Value of $\phi$ for $b/b_0=10^{-2}$. The vertical curves correspond to $\epsilon/b_0=2^{1/\ga}$.}
\label{repulsive_eps}
  \end{center}
\end{figure}

\begin{figure}
  \begin{center}
    \psfrag{X}{$b/b_0$}
    \psfrag{Y}[c]{$\phi_\epsilon(b,b_0)/\phi_0(b/b_0)-1$}
    	\psfrag{AAAAAAAA}[c][c]{$\gamma=1/2$}
  	\psfrag{BBBBBBBB}[c][c]{$\gamma=1$}
  	\psfrag{CCCCCCCC}[c][c]{$\gamma=3/2$}
   	{\includegraphics[height= 0.35\textwidth]{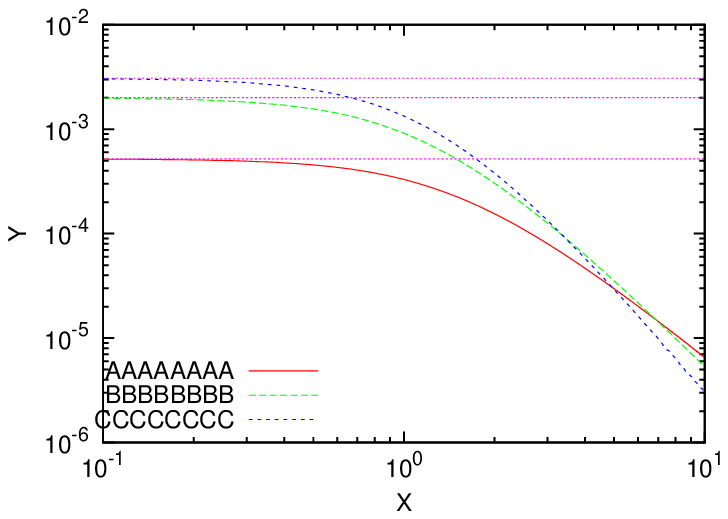}}\\
   	\psfrag{X}{$b/b_0$}
    \psfrag{Y}[c]{$\pi/2-\phi_\ep(b,b_0)$}
    {\includegraphics[height= 0.35\textwidth]{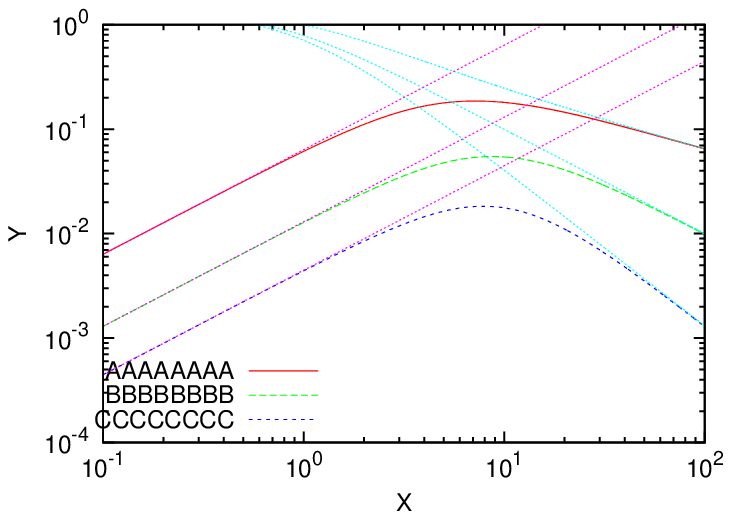}}
\caption{ Numerical computations for repulsive potentials with Plummer softening. 
Top: Graph of $\phi_\ep$ normalized to the angle without softening $\phi_0$(continuous line) 
and of the leading order term (dotted line) given in Eq.~\eqref{hardrepulssoft} 
as a function of $b / b_0$ for different values of $\gamma$ and $\epsilon/b_0=1/10$.
Bottom: Graph of $\phi_\ep$ for $\epsilon/b_0=10$ and the leading order expansion given in 
Eq.~\eqref{hardrepulssoft-autre}. The dotted blue lines corresponds to  $\phi_0$.
}
\label{repulsive_softening}
  \end{center}
\end{figure}

In the case of repulsive interactions, we have seen that two different
behaviours are predicted depending whether $\epsilon/b_0$ is larger
than $2^{1/\ga}$ or not. In the case $\epsilon/b_0<2^{1/\ga}$, the
softening does not modify strongly the angle $\phi$: it behaves
linearly for $b\ll b_0$, only its slope is modified with $\epsilon$.
In the case in which $\epsilon/b_0>2^{1/\ga}$, hard collisions are
radically modified, obtaining $\lim_{b/b_0\to0} \phi=\pi/2$.  The
change of behaviour occurs sharply at $\epsilon/b_0=2^{1/\ga}$ as we
show in Fig.~\ref{repulsive_eps}, in which $\phi$ is plotted as a
function of $\epsilon$ at fixed $b$, for some values of $\ga$. The
range of validity in $b$ of the linear correction is given by the {\it
  largest} value of $b_0$ and $\epsilon$.
In Fig.~\ref{repulsive_softening} we show the comparison between the numerical 
integration of $ \phi_\ep $ in Eq.~\eqref{expressphisoft} 
with the asymptotic predictions Eqs.~\eqref{hardrepulssoft} and \eqref{hardrepulssoft-autre}. 
We see a very good  matching between the curves.   

\begin{figure}
  \begin{center}
    \psfrag{X}[c][c]{$b/b_0$}
    \psfrag{Y}[c][c]{$\phi_\epsilon(b,b_0)-\pi/2$}
    	\psfrag{AAAAAAAA}[c][c]{$\gamma=1/2$}
  	\psfrag{BBBBBBBB}[c][c]{$\gamma=1$}
  	\psfrag{CCCCCCCC}[c][c]{$\gamma=3/2$}
   	{\includegraphics[height= 0.35\textwidth]{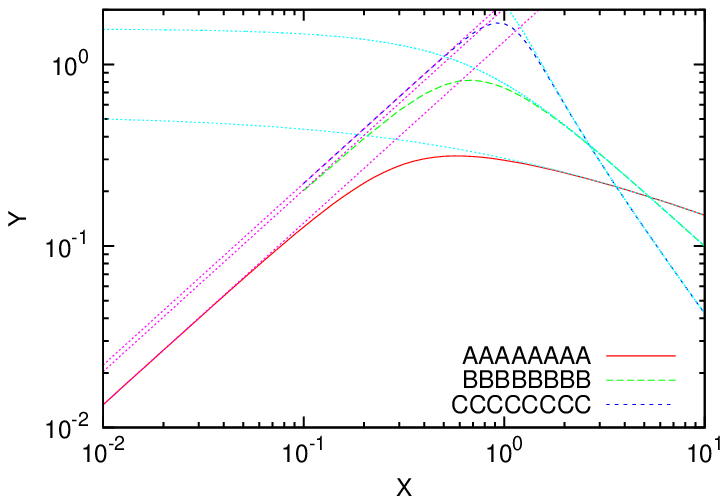}}\\
    {\includegraphics[height= 0.35\textwidth]{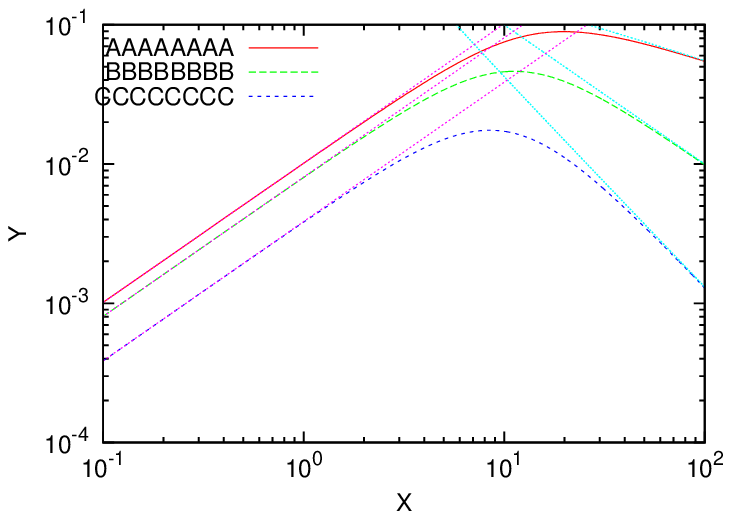}}
\caption{Numerical computations for attractive potentials with Plummer softening 
(hard scattering).
Top: Graphs of $\phi_\ep$ (continuous line) and the theoretical prediction 
Eq.~\eqref{hardrepulssoft} (dotted lines) as a function of $b / b_0 $ for different 
values of $\gamma$ and $\epsilon/b_0=1/10$.  
Bottom: same quantity for $\epsilon/b_0=10 $ and the theoretical prediction 
Eq.~\eqref{hardrepulssoft-autre} (dotted lines).  The dotted blue lines corresponds to  $\phi_0$.} 
\label{attractive_softening}
  \end{center}
\end{figure}

\begin{figure}
  \begin{center}
  \psfrag{AAAAAA}[c][c]{$\epsilon>\ep_*$}
  \psfrag{BBBBBB}[c][c]{$\epsilon<\ep_*$}
    \psfrag{X}{$b/b_0$}
    \psfrag{Y}{$\phi_\epsilon(b,b_0)$}
       {\includegraphics[height= 0.35\textwidth]{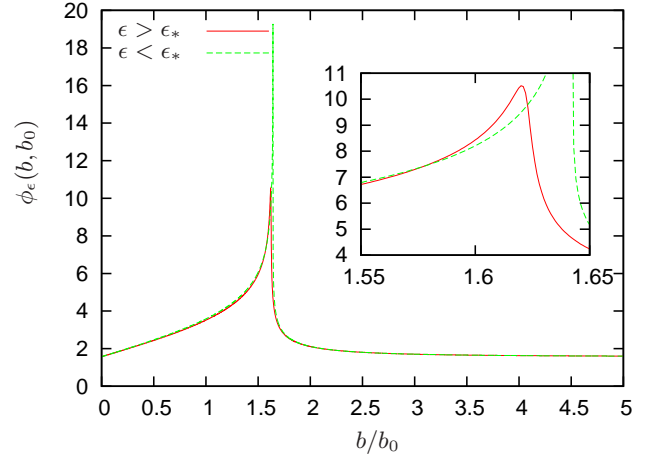}}
\caption{Plot of $\phi_\ep$ as a function of $b$ for $\gamma=5/2$ and two different values 
of the softening. 
The red continuous curve corresponds to a value of $\epsilon$ slightly larger than 
$\ep_*(b_0, \ga)$ and the dashed green one to a value of $\epsilon$ slightly smaller 
than $\ep_*(b_0, \ga)$.}
\label{gamma2.5eps}
  \end{center}
\end{figure}

For the case of attractive interactions, the range of validity in $b$ of the linear correction is always given by $\ep$. In Fig.~\ref{attractive_softening} we show a very good agreement matching between the exact integration  Eq.~\eqref{expressphisoft} with the asymptotic predictions Eqs.~\eqref{hardrepulssoft} and \eqref{attractifhard-soft-autre}.
We have also studied, for $\gamma>2$, for which value of the softening, there is no formation of pairs for any value of $b$. We have seen that  introducing a softening $\epsilon>0$  automatically regularizes the angle $\phi$ for any value $b$, except one, for which there is {\it orbiting}, except for some particular softenings, in which the divergences also disappear. It is necessary to introduce a value of the softening larger than a critical value (which we have calculated explicitly) to regularize completely the problem. In Fig.~\ref{gamma2.5eps} we illustrate this behavior. The continuous red curve corresponds 
to the case in which $\epsilon>\ep_*(b_0, \ga) $. In this case, $\phi_\ep$ is a regular 
function of $b$, as it can be seen in the inset. The dashed green curve corresponds to the 
case in which $\epsilon<\ep_*(b_0, \ga) $, for which $\phi_\ep$ diverges for $b=b_\sharp(\ep)$. 

\section{Conclusions}

In this paper we have studied the scattering of two particles
interacting with a central potential $v(r)\sim 1/r^\ga$. This is a
generalization of the Rutherford formula of the scattering of two
particles interacting via a Coulomb or gravitational force. Unlike the
original case, it is not possible to compute in general the deflection
angle of the particles analytically for general $\gamma\ne1$.  We have
then calculated the asymptotics of the angle of deflection for the two
limiting cases in which we are interested in: the {\it weak}
collisions regime, in which the particles trajectories are weakly
perturbed, and the {\it strong} collision regime, in which they are
strongly perturbed.  Combining the analytical expressions and the
numerical integration of the equation of motion, we have derived the
phenomenology we detail as follows.

In the regime of soft collisions, attractive and repulsive interactions give a very 
similar result: the angle of closest approach scales as 
\be
\phi\sim \pi/2 \mp A(\ga) \left(\frac{b}{b_0}\right)^\gamma,
\ee
where $A(\ga)>0$, $b$ is the impact factor, and $b_0$ a characteristic
scale which depends on the reduced mass, the coupling constant and the
relative asymptotic velocity of the particles. The minus sign
corresponds to the repulsive interaction and the positive sign to
attractive interactions. This is what it is expected: with no
interaction, the reduced particle suffers no deflection, and then
$\phi = \pi/2$. If the interaction is repulsive, the reduced particle
will be deflected in the top left quadrant, which implies that $\phi <
\pi/2$. If the interaction is attractive, it will be deflected to the
bottom left quadrant, which implies $\phi > \pi/2$ (see
Fig.~\ref{coll}).

 In the regime of hard collisions, the situation is very different between the repulsive 
 and attractive case: {\it(i)} for repulsive interactions, the angle of closest approach 
 scales as $\phi\sim b/b_0$. This is what one expects for $b\to0$: for vanishing impact 
 factor, particle bounce one on each other, coming back in their original directions 
 and opposite sign of the velocity;  {\it(ii)} for attractive interactions 
 with $\ga < 2 $, the leading contribution is 
\be
\label{number-of-loops} 
\phi\sim \frac{\pi}{2-\ga}.
\ee
We see therefore, that for $b/b_0\to0$ the angle of deflection depends on the exponent 
of the  interaction potential $\ga$. 
Of course, the deflection angle is the same for Coulomb (repulsive) 
and gravitational interaction ($\gamma=1$)

Equation \eqref{number-of-loops} implies that, when $\ga$ approaches the value of $2$, the 
angle $\phi$ increases, diverging in the limit $\ga\to 2$. This is due to the effective 
potential created by the angular momentum, which scales with the distance as $1/r^2$. 
 When the exponent $\ga$ of the attractive potential is larger than $2$, 
the angular momentum term cannot, in general, prevent the system to collapse and the particles 
crash.
Studying the distance of closest approach $r_{min}$ we have found two different behaviors 
whether $\ga$ is smaller or larger than $2$:

\begin{itemize}

\item If $\ga<2$, in the limit  $\ga\to2^-$ (for any $b$ smaller than some critical value which we have calculated explicitly), the value of 
$r_{min} $ tends to $0$. 
The trajectories in this limit is a succession of smaller and smaller loops embedded 
one in the other. An example of such trajectories was given in Fig.~\ref{coll2}.

\item If $\ga>2$, the particles do not crash if the impact factor is 
larger than some critical value, which we have calculated. For impact factor  slightly 
larger than 
this critical value, we have trajectories with 
$r_{min}\sim b_0$. The particles then {\it orbite} with distance $r_{min}$ forming a binary, which will be destroyed in a finite time. We gave an example 
of such trajectories in Fig.~\ref{gamma2.05}.

\end{itemize}

We have also studied the effect of introducing a regularization at small scales in the potential. The conclusions are detailed in Subsect.~\ref{summary-soft}. 

A practical  application appears naturally in the context of astrophysics or plasma physics, when  we are interested in calculating the average change of velocity due to the collisions.
It is usual  (see e.g. \cite{binney}) to decompose the relative velocity of the particles before the collisions  $\bV$ as the sum of its component  
along the direction of the initial relative velocity ${\mathbf e}_{\parallel}$ and the
component perpendicular to it ${\mathbf e}_{\perp}$, i.e.,
\be
\bV=V_{\perp} {\mathbf e}_{\perp} +  V_{\parallel} {\mathbf e}_{\parallel}.
\ee
It is possible to compute the average change of
velocity $\Delta V_\perp$ and $\Delta V_{\|}$  after a collision has
been completed  integrating over all the impact factors $b$:
\bse
\label{v_perp-and-v_par}
\begin{align}
\frac{\Delta V_\perp}{V} &= \sin(2\phi)\\
\frac{\Delta V_\parallel}{V} &= 1+\cos(2\phi).
\end{align}
\ese
One quantity of interest is the average change velocity square, which can be expressed by the integral over all the impact factors, i.e.,
\bse
\begin{align}
\label{total-change-v}
\langle\Delta V_\perp^2\rangle &\sim \int_{0}^{R} db b ^{d-2}\sin^2\left(2\phi_\ep\left(\frac{b}{b_0}\right)\right)\\
\langle\Delta V_\parallel^2\rangle &\sim \int_{0}^{R} db b^{d-2}\left[1+\cos\left(2\phi_\ep\left(\frac{b}{b_0}\right)\right)\right]^2,
\end{align}
\ese
where $d>1$ is the physical dimension and $R$ the size of the system, which is the maximal impact factor available.

In astrophysical or cosmological N-body simulations, the goal is to simulate {\it collisionless} dynamics sampling a continuous distribution with macro-particles (see e.g. \cite{athanassoula_00}).  The softening used in these simulations is much larger than $b_0$ (in order to suppress collisional effects), and hence (see Sect.~\ref{theo-soft}), $\phi-\pi/2\ll 1$. We can therefore write 
\be
\label{total-change-v2}
\langle\Delta V_\perp^2\rangle \sim 4\int_{0}^{R} db b^{d-2}\left[\phi_\ep\left(\frac{b}{b_0}\right)-\frac{\pi}{2}\right]^2
\ee
and $\langle\Delta V_\parallel^2\rangle\ll \langle\Delta V_\perp^2\rangle$. We can estimate Eq.~\eqref{total-change-v2} using the following approximate expression for the angle $\phi_\ep$ (we will consider explicitly  attractive interactions with Plummer softening to simplify notations, the compact softening or repulsive case is analogous):
\be
\label{phi-approx-large-eps}
\phi_\ep-\frac{\pi}{2} \simeq \left\{ 
\begin{array}{ll}
C_\ep(\gamma)\frac{b}{\ep} & \text{if } b < \ep \\
 A(\ga)\left(\frac{b_0}{b}\right)^\ga & \text{if } b > \ep
\end{array}
\right.
\ee
(see e.g. Fig.~\ref{attractive_softening}). Using Eq.~\eqref{phi-approx-large-eps} to compute integral \eqref{total-change-v2}, considering softenings such that $b_0\ll\ep\ll R$ we get the scaling, for $\ga>(d-1)/2$,
\be
\label{total-change-v-gne1}
\langle\Delta V_\perp^2\rangle\sim b_0^{2\ga} \epsilon^{d-1-2\ga}
\ee
where we have used the asymptotic value of $C_\ep(\ga)$ Eq.~\eqref{C-large-softening}.  Notice that impact factors smaller or larger than $\ep$ contributes to the scaling \eqref{total-change-v-gne1}. In the limiting case $\ga=(d-1)/2$, we get
\be
\langle\Delta V_\perp^2\rangle\sim b_0^{2} \ln\left(\frac{R}{\epsilon}\right).
\ee
In this case contributions of collisions with $b<\epsilon$ are negligible. For $\ga<(d-1)/2$, the effect  of the softening is negligible because the main contribution to the change of velocity is given by impact factors $b\sim R$.

\section*{Acknowledgments}

We thank M.~Joyce for useful discussions and comments.

\appendix

\section{Mathematical details}

In this appendix we give mathematical details of some derivations given in the paper.

Some of the integrals appearing in the paper may be expressed with the help of the Beta function 
(also called Euler's integral of the first kind) defined for $x $, $ y>0 $ by
\begin{align*} 
{\mathbb B} (x, y) 
&  = \int_0^1 t^{x-1} (1-t)^{ y -1 } \, dt \\
& = 2 \int_0^{\pi/2} \sin^{2x -1} (\vartheta ) \cos^{2y-1} (\vartheta ) \, d \vartheta\\
& = \frac{ \Gamma (x) \Gamma(y) }{\Gamma( x+y)} ,
\end{align*}
where $ \Gamma $ is Euler's function.

\subsection{Expression for the integral Eq. \eqref{integrale1}}
\label{app1}

For the integral Eq.~\eqref{integrale1}, we use the substitution $ x = \cos \vartheta $ and integration 
by parts:
\begin{align}
\label{integrals} 
\nonumber
& \int_0^1 \frac{ 1 - x^\ga}{ ( 1 -x^2 )^{3/2}} \, dx 
= \int_0^{\pi /2 } \frac{1 - \cos^\ga ( \vartheta) }{ \sin^2 \vartheta } \, d \vartheta \\
& =  
\left[ \frac{\cos^\ga ( \vartheta) - 1 }{\tan \vartheta } \right]_0^{\pi /2 }
+ \ga \int_0^{\pi /2 } \cos^\ga ( \vartheta) \, d \vartheta .
\end{align}
Notice that the bracket term vanishes. The right-hand side 
of Eq.~\eqref{integrals} may also be expressed 
as (using that $ \Gamma ( 1 + z ) = z \Gamma (z) $)
\be
\label{beta-gamma}
  \ga {\mathbb B} \left( \frac{ \ga +1 }{2} , \frac12 \right) 
 = \sqrt{\pi} \frac{ \Gamma \left( \frac{ \ga +1 }{2} \right) }{\Gamma \left( \frac{\ga}{2} \right)} .
\ee

\subsection{Extension of the expansion  Eq.~\eqref{flowerpower} for $ \gamma \in ( 0 , 2 ] $}
\label{appextension}

We proceed in two steps: we first prove that $ \phi $ is a power series in $ ( b_0 / b )^\gamma $ for $b$ sufficiently large, and then identify the coefficients in the expansion.

The argument used for Eq.~\eqref{prototype} shows that $ (r_{min}/b) \phi $ is a power series of the variable $ \delta $ (with positive radius) provided $ \delta $ is small enough. Moreover, since
$$
 \frac{b}{r_{min}}
= \sqrt{ 1 \pm 2 ( b_0 / r_{min} )^\ga }
= \sqrt{ 1 \pm 2 ( b_0 / b )^\ga ( b / r_{min} )^\ga } ,
$$
it is easy to show that $ b / r_{min} $, thus also $\delta= \pm ( ( b / r_{min} )^2 - 1 ) $, is itself a power series of the variable $ 2 ( b_0 / b )^\ga $ (with positive radius). By substitution and Cauchy product, $ \phi $ is a power series
in $ 2 ( b_0 / b )^\gamma $ for $b$ sufficiently large, that is there exists some coefficients $\kappa_n ( \gamma ) $, $n \in \mathbb{N} $, such that, for $b$ large enough,
$$
 \phi = \sum_{n=0}^{+\infty} \kappa_n (\gamma ) \left( 2 ( b_0 / b )^{\gamma } \right)^n .
$$
In addition, from the above computation, we know that each coefficient $ \kappa_n ( \gamma ) $ is a finite sum of the type
$$
\sum_{k=0}^n C(n,k) \int_0^1 \left( \frac{ x^\ga - x^2 }{1- x^2} \right)^k \frac{dx}{ \sqrt{ 1-x^2} } ,
$$
the integrals coming from the expansion of the integral $ (r_{min}/b) \phi $ in powers of $ \delta $, and the coefficients $C(n,k) $ of the Cauchy products and the substitution.
In particular, each coefficient $ \kappa_n ( \gamma ) $ is an analytic function of $\gamma $ in $ ( 0, +\infty ) $ (and even in the half-space $ \{ {\rm Re} >0 \} $).

We now identify the coefficients $ \kappa_n ( \gamma ) $ by considering the two expansions valid for $ \gamma > 2 $ and $b$ large,
\begin{align*}
 \phi
& = \sqrt{\pi} \sum_{n= 0}^{+\infty }
\frac{ \Gamma ( (n \gamma +1) /2 ) }{ 2 n! \Gamma ( 1 + n ( \gamma /2 - 1 ) ) } ( \mp 2 ( b_0/ b)^\gamma)^n
\\
& = \sum_{n=0}^{+\infty} \kappa_n (\gamma ) \left( 2 ( b_0 / b )^{\gamma } \right)^n .
\end{align*}
By uniqueness of the power series expansions, we deduce that if $ \gamma > 2 $, then for all $ n \in \mathbb{N} $,
$$
 \kappa_n (\gamma )
= ( \mp 1)^n \sqrt{\pi} \frac{ \Gamma ( (n \gamma +1) /2 ) }{ 2 n! \Gamma ( 1 + n ( \gamma /2 - 1 ) ) } .
$$
Since $ \kappa_n $ is an analytic function in $ ( 0, +\infty ) $ and both
$ \gamma \mapsto \Gamma ( (n \gamma +1) /2 ) $ and
$ \gamma \mapsto 1 / \Gamma ( 1 + n ( \gamma /2 - 1 ) ) $ are analytic in $ ( 0, +\infty ) $,
we deduce from the principle of permanence for analytic functions that
Eq.~\eqref{flowerpower} holds true for any $ \gamma > 0 $.

We may now compute the radius of convergence. If $ \gamma > 2 $, this has been carried out
in \cite{mott-smith_60} using the generalized Stirling formula
$ \Gamma ( s+1 ) \approx ( s/{\sf e} )^s \sqrt{2\pi s } $ when $s \to + \infty $.
The generalization to $\gamma \leq 2 $ follows from the same type of computations,
combined with Euler's reflection formula $ \Gamma (s) \Gamma ( 1 - s ) = \pi / \sin( \pi s ) $. 

\subsection{Expression for the integral Eq. \eqref{integrale2}}
\label{app2}

Using the substitution $ x^\ga = \cos^2 \vartheta $ provides
\begin{align*}
 \int_0^1 \frac{dx}{ \sqrt{ 1 - x^\ga } } 
 & = \frac{2}{\gamma} \int_0^{\pi/2} \cos^{ \frac{2}{\gamma} - 1 } \vartheta \, d \vartheta 
 \\ 
 & = \frac{1}{\ga } {\mathbb B} \left( \frac{1}{\ga} , \frac12 \right) 
 = \frac{ \sqrt{\pi } \Gamma\left(1+\frac{1}{\ga}\right)}{ 
 \Gamma\left(\frac{1}{2}+\frac{1}{\ga}\right)} ,
\end{align*}
as claimed.

\subsection{Expression for the integral Eq. \eqref{integrale3}}
\label{app3}

In Eq. \eqref{integrale3}, we substitute $ x^{2 - \ga} = \cos^2 ( \vartheta ) $ 
and then integrate by parts
\begin{align*}
 \int_0^1 & \frac{ 1 - x^\ga }{ 2 ( x^\ga - x^2 )^{3/2} } \, dx 
\\ & = \frac{1}{ 2 - \ga } 
 \int_0^{\pi /2} \frac{\cos^{-2 \ga/ ( 2- \ga)} (\vartheta) -1}{\sin^2 \vartheta}\, d \vartheta 
 \\ 
& = \frac{1}{ 2 - \ga} \left[ \frac{\cos^{-2 \ga/ ( 2- \ga)} ( \vartheta) - 1 }{\tan \vartheta } \right]_0^{\pi /2} 
 \\ & \quad + 
 \frac{2 \ga}{ (2 - \ga)^2 } \int_0^{\pi /2} \cos^{-2 \ga/ ( 2- \ga)} (\vartheta) \, d \vartheta.
\end{align*}
Since the bracket vanishes, we then obtain, as wished,
\begin{align*}
\int_0^1 \frac{ 1 - x^\ga }{ 2 ( x^\ga - x^2 )^{3/2} } \, dx 
& = 
\frac{ \ga}{ (2 -\ga)^2 } {\mathbb B} \left( \frac{2 - 3 \ga}{2(2-\ga)}, \frac12 \right) 
\\
& = 
 \frac{ \ga}{ (2 -\ga)^2 } 
 \frac{\sqrt{\pi} \Gamma \left(\frac{2 - 3 \ga}{2(2-\ga)} \right) }{ 
 \Gamma \left(\frac{2 (1- \ga)}{2-\ga} \right) } .
\end{align*}

\subsection{Expression for the integral Eq.~\eqref{integrale4}}
\label{app4}

In the integral $ \int_0^{ + \infty } \Psi_0 (y) \, dy $, we use successively 
the substitutions $ y^\ga = \sinh^2 u $ and $ {\sf e}^{-u} = \sin \vartheta $:
\begin{align*}
\int_0^{ + \infty } \Psi_0 (y) \, dy 
& = \int_0^{+\infty} \frac{ y^{ - \ga/2 } ( 1 + y^\ga )^{-1/2 } dy}{y^{ \ga /2} + ( 1 + y^\ga )^{1/2 } } 
\\
& = \frac{2}{\ga} \int_0^{+\infty} (\sinh u)^{ \frac{2}{\ga} -2 } {\sf e}^{-u} \, du 
\\ 
& = \frac{2^{3- 2 /\ga}}{\ga} \int_0^{\pi / 2} (\sin \vartheta )^{ 2 - \frac{2}{\ga} } 
(\cos \vartheta )^{ \frac{4}{\ga} - 3} \, d \vartheta
\\ & = \frac{2^{3 - 2 /\ga}}{\ga} \mathbb{B} \left( \frac32 - \frac{1}{\ga} , \frac{1}{\ga}  - 1 \right) 
\\ & = 
\frac{2^{3 - 2 /\ga}}{\ga} \frac{ \Gamma \left( \frac32 - \frac{1}{\ga} \right) 
\Gamma \left( \frac{2}{\ga } -1 \right)}{\Gamma \left( \frac{1}{\ga } + \frac12 \right)} ,
\end{align*}
which establishes the equality Eq. \eqref{integrale4}.

\subsection{Justification of Eq.~\eqref{guess}}
\label{app30}
We may already get rid of the  contribution for $ 0 \leq y \leq 1 $ since we know that $ Q (\de ) \to +\infty $ 
whereas
$$
 \int_0^1 \Psi_\de (y) \, dy \to \int_0^1 \Psi_0 (y) \, dy < \infty ,
$$
since the integrand is $ \approx y^{- 1/3} $ at the origin. Then, Eq. \eqref{mortel} implies
$$
 Q ( \de ) \approx \int_1^{ \de^{-3/2} / 2 } \Psi_\de (y) \, dy .
$$
To prove the asymptotics Eq.~\eqref{guess} rigorously, we have to pay attention to the $y$'s 
close to $ \de^{-3/2} / 2 $. Indeed, when $ \ga = 2/3 $,
$$ 
( y^\ga - \de^{2/\ga -1} y^2 )^{-1/2} = y^{- 1/3} ( 1 - \de^{2} y^{4/3} )^{-1/2} 
$$
but $ \de^{2} y^{4/3} $ may be of order one when $y \approx \de^{-3/2} $. 
Therefore, we split
\begin{align*}
 Q ( \de ) \approx Q_1(\de) + Q_2 (\de ) & = \int_0^{\de^{-3/2} / \lvert \ln \de \rvert } \Psi_\de (y) \, dy 
\\ & \quad + \int_{\de^{-3/2}/ \lvert \ln \de \rvert}^{\de^{-3/2}/2 } \Psi_\de (y) \, dy .
\end{align*}
For $ Q_1 (\de ) $, we may write, factorizing the dominant terms,
\begin{align}
\label{factor}
( y^\ga - \de^{2/\ga -1} y^2 )^{-1/2} 
& = y^{-1/3} ( 1 - \de^2 y^{4/3} )^{-1/2} 
\nonumber \\
& = y^{-1/3} ( 1 + o(1) ) ,
\end{align}
since $ 0 \le \de^2 y^{4/3} \le \lvert \ln \de \rvert^{-1} = o(1) $ and 
where $ o(1) $ stands for a quantity which is uniformly small 
for $ y \in [ 0 , \de^{-3/2} / \lvert \ln \de \rvert ] $. Similarly 
$ 1 - \de y^\gamma = 1 - \de y^{2/3} = 1 + o(1) $ and 
\begin{align*}
& ( y^\ga - \de^{2/\ga -1} y^2 + 1 - \de y^\ga )^{-1/2} 
\\
& = (y^{2/3} +1)^{-1/2} \left( 1-\frac{\de^2 y^{4/3} + \de y^{2/3} }{1 + y^{2/3} } \right)^{-1/2} \\
& = (y^{2/3} +1)^{-1/2} ( 1 + o(1) ) 
\end{align*}
and we then infer
\begin{align*}
Q_1 (\de ) & = \int_1^{\de^{-3/2} / \lvert \ln \de \rvert } ( 1 +o(1) ) 
\frac{ y^{-1/3} ( y^{2/3} + 1 )^{-1/2}\, dy}{y^{1/3 } + \sqrt{ y^{2/3} + 1 } } 
\\ 
& \approx 
\int_1^{\de^{-3/2} / \lvert \ln \de \rvert } \frac{dy}{2 y } 
= 
\frac12 \ln \left( \de^{-3/2} / \lvert \ln \de \rvert \right) \\ 
& \approx \frac34 \lvert \ln \de \rvert,
\end{align*}
since the first integrand is $ \approx 1/(2y) $ at infinity. 

We now consider $ Q_2 (\de ) $, where 
$ 1 \ll \de^{-3/2} / \lvert \ln \de \rvert \le y \le \de^{-3/2} /2 $. 
Then, in Eq.~\eqref{factor}, we no longer have a $ o(1) $, but we can write, 
since $ 0 \le \de^2 y^{4/3} \le 1/2 $,
$$
( y^\ga - \de^{2/\ga -1} y^2 )^{-1/2} 
= y^{-1/3} ( 1 - \de^2 y^{4/3} )^{-1/2} = \mathcal{O}( y^{-1/3} ) 
$$
and similarly
$$
( y^\ga - \de^{2/\ga -1} y^2 + 1 - \de y^\ga )^{-1/2} = \mathcal{O} (y^{-1/3} ) .
$$
As a consequence, 
$$
 Q_2 (\de ) = \mathcal{O} \left( \int_{\de^{-3/2} / \lvert \ln \de \rvert}^{\de^{-3/2} /2 } 
 \frac{dy}{y} \right) = \mathcal{O} \left( \ln ( \lvert \ln \de \rvert ) \right) 
 \ll \lvert \ln \de \rvert .
$$
Combining the estimates for $ Q_1 (\de )$ and $ Q_2 ( \de ) $, we have justified Eq.~\eqref{guess}.

\subsection{Justification of the leading order expansion Eq. \eqref{integrale10}}
\label{app10}

To completely justify the expansion Eq.~\eqref{integrale10}, we have to pay attention to the $z$'s 
close to $ z_{max} $. Notice first that
$$ 
 dr /dz = \sqrt{ - 2 W_b (r_*) / W_b''(r_*)} ( 1 + \mathcal{O} ( z / z_{max} ) ) 
$$
and that
$$ 
 r (z)^{-2} = ( r_* + \mathcal{O}  ( z / z_{max} ) )^{-2} ,
$$
hence the asymptotics $ r (z) \approx r_* \approx R $ and 
$ dr / dz \approx \sqrt{ - 2 W_b (r_*) / W_b''(r_*)} $ are not completely true for $z \sim z_{max} $. 
We therefore split the right-hand side of Eq. \eqref{equivalent} as
\begin{align*}
I_1 + I_2 = \frac{b}{ \sqrt{ - W_b (r_*) } } \int_1^{z_{max} / \ln( z_{max} ) }
 \frac{ r (z)^{-2} \, dr / dz }{  \sqrt{ z^2 - 1 } } \, dz 
\\ 
+ \frac{b}{ \sqrt{ - W_b (r_*) } } \int_{z_{max} / \ln( z_{max} ) }^{ z_{max} } 
 \frac{ r (z)^{-2} \, dr / dz }{  \sqrt{ z^2 - 1 } } \, dz .
\end{align*}
In $ I_1 $, we have $ 0\leq z / z_{max} \leq \lvert \ln z_{max}\rvert = o(1) $, thus
$$ 
dr /dz = \sqrt{ - 2 W_b (r_*) / W_b''(r_*)} ( 1 + o(1) ) 
$$
and
$$ 
r (z)^{-2} = ( r_* + o(1) )^{-2} = R^{-2} + o(1) ,
$$
which yields
\begin{align*}
I_1 & \approx b \sqrt{ \frac{ 2}{ W_b''(R)} } \int_1^{z_{max} / \ln( z_{max} ) }
 \frac{ R^{-2} \, dz }{  \sqrt{ z^2 - 1 } } 
 \\
 & \approx \sqrt{ \frac{ 2}{ R^{4} W_b''(R)} } \ln( z_{max} ) .
\end{align*}
Turning back to $ I_2 $, where $ 1\ll z_{max} / \ln( z_{max} ) \leq z \leq z_{max} $, we simply use that 
$ r(z)^{-2} = \mathcal{O} ( 1 ) $ and that $ dr /dz  = \sqrt{ - 2 W_b (r_*) } \mathcal{O} (1) $, thus
\begin{align*}
I_2 & = \mathcal{O} \left( \int_{z_{max} / \ln( z_{max} ) }^{ z_{max} } \frac{dz}{z} \right) 
\\ 
& = \mathcal{O} ( \ln (\ln z_{max} ) ) \ll \ln ( z_{max} ) .
\end{align*}
This concludes the justification of Eq. \eqref{integrale10}.

\subsection{Bounding the function $ \dfrac{1-x^2}{ F (x, r_{min}/\ep) } $ }
\label{appbornage1}

We prove here that the function $ x \mapsto \frac{1-x^2}{ F(x, r_{min} / \ep) } $ 
is bounded on $[0,1] $, independently of $b \ll b_0 $ (for the Plummer softening). 
We recall that for the regime ($ \ep < b_0 2^{1/\ga} $ and $ b \ll b_0 $) 
we are studying, $ r_{min} \approx b_0 2^{1/\ga} \sqrt{ 1- \hat{\ep}^2 } $, thus 
$ r_{min} / \ep \approx \hat{\ep}^{-1} \sqrt{ 1- \hat{\ep}^2 } $. 

Let us first work on the interval $ [ 0 , 1/2 ] $. Then, 
$ F (x,r_{min}/\ep) = 1 - \mathcal{V}^{\text{Pl}} (r_{min} / (\ep x) ) /\mathcal{V}^{\text{Pl}} (r_{min} /\ep ) $ 
is decreasing with respect to $x$ since $ \mathcal{V}^{\text{Pl}} (R) = ( 1 +R^2)^{-\ga/2} $ 
is decreasing on $ [ 0, + \infty ) $, hence, for $ 0 \leq x \leq 1 / 2 $,
$$
0 \leq \frac{1-x^2}{F(x,r_{min} /\ep ) } 
 \leq \frac{1}{ F(x, r_{min} / \ep) } 
 \leq \frac{1}{ F (1 / 2 , r_{min}/\ep) } .
$$
The right-hand side does not depend on $x$ and is equal to
$$ 
 \left( 1 - \frac{\mathcal{V}^{\text{Pl}} ( 2 r_{min}/\ep )}{ \mathcal{V}^{\text{Pl}} ( r_{min}/\ep) } \right)^{-1} 
 \approx 
 \left( 1 - \frac{\mathcal{V}^{\text{Pl}}( 2 \hat{\ep}^{-1} \sqrt{1 - \hat{\ep}^2})}{ \mathcal{V}^{\text{Pl}} (\hat{\ep}^{-1} \sqrt{1 - \hat{\ep}^2} ) } \right)^{-1} ,
$$
which gives the desired upper bound on $ [ 0 , 1 / 2 ] $.

We now work on $ [ 1 / 2 , 1 ] $, and use that 
$ \frac{d}{dx} \mathcal{V}^{\text{Pl}} ( r_{min}/ ( \ep x) ) 
= - ( r_{min}/ ( \ep x^2 ) ) ( \mathcal{V}^{\text{Pl}} )' ( r_{min}/ ( \ep x) ) \geq m $ 
for some positive constant $m = m (\hat{\ep}) $ independent of $b$, since $ \mathcal{V}^{\text{Pl}} $ 
is decreasing on $ [ 0, + \infty ) $. As a consequence of the mean value theorem we get
$$
 0 \leq \frac{1-x^2}{F(x,r_{min} / \ep) } = \frac{(1+x)(1-x)}{F(x,r_{min} /\ep) - F(1,r_{min} /\ep ) } 
 \leq \frac{2}{m} . 
$$
This concludes the proof of the upper bound on $ [ 0 , 1 / 2 ] $.

\subsection{Justification of the relation Eq. \eqref{integrale14}}
\label{app14}

If $ \hat{\ep } \gg 1 $, we may use for instance the Taylor expansion of the square root 
to deduce
\begin{align*}
\label{G-eps-large}
\tilde{B}_{\hat{\ep}} ( \ga ) & \approx 
\int_0^{+\infty} \left( 1 - \frac{1}{\sqrt{ 1 + \frac{1}{ \hat{\ep}^{\ga} -1} 
\left( \mathcal{V}^{\text{Pl}}(0) - \mathcal{V}^{\text{Pl}}(y) \right)} } \right) \, \frac{dy}{y^2}
\\ 
& \approx \frac{1}{ \hat{\ep}^{\ga} -1} \int_0^{+\infty} 
\frac{ \mathcal{V}^{\text{Pl}}(0) - \mathcal{V}^{\text{Pl}}(y) }{ 2 y^2 } \, dy 
\\ 
& \approx \frac{1}{ \hat{\ep}^{\ga} } \int_0^{+\infty} \frac{ 1 - ( 1 + y^2 )^{- \ga/2} }{ 2 y^2 } \, dy 
\nonumber \\ & =  \frac{\ga}{4} \int_0^{+\infty} ( 1 + y^2 )^{- \ga/2 - 1} \, dy 
\nonumber \\ & = \frac{\ga}{4} \int_0^{ \pi /2 } \cos^\ga ( \vartheta ) \, d \vartheta 
= \sqrt{\pi} \frac{ \Gamma \left( \frac{ \ga +1 }{2} \right) }{ 4 \Gamma \left( \frac{\ga}{2} \right)} ,
\end{align*}
by first integration by parts and then the use of the substitution $ y= \tan \vartheta $.

\end{document}